\documentclass[preprint2]{aastex}

\usepackage{subfigure,lscape}
\input{psfig.sty}

\shorttitle{Deep ROSAT-HRI observation of of the NGC 1399/NGC 1404 region}
\shortauthors{Paolillo et al.}

\begin{document}

\title{Deep ROSAT-HRI observations of the NGC 1399/NGC 1404 region: morphology and structure of the X-ray halo}
\author{M. Paolillo\altaffilmark{1}, G. Fabbiano\altaffilmark{2}, G. Peres\altaffilmark{1}, D.-W. Kim\altaffilmark{2}\\
\centerline{\it 25 June 2001}}

\altaffiltext{1}{Universit\`a di Palermo - DSFA - Sez.di Astronomia, P.zza del Parlamento 1, 90134 Palermo; paolillo@astropa.unipa.it, peres@astropa.unipa.it}
\altaffiltext{2}{Harvard-Smithsonian Center for Astrophysics, High Energy Division, 60 Garden St., Cambridge, MA 02138; pepi@head-cfa.harvard.edu, kim@head-cfa.harvard.edu}

\begin{abstract}

We present the analysis of a deep (167 ks) ROSAT HRI observation of the cD galaxy NGC 1399 in the Fornax cluster, comparing it with previous work on this galaxy and with recent {\it Chandra} data. We find, in agreement with previous observations, an extended and asymmetric gaseous halo with a luminosity (in the 0.1-2.4 keV energy band) of $L_X=(5.50\pm 0.04)\times10^{41}$ erg s$^{-1}$ within 46 kpc (assuming a distance of D=19 Mpc). Using both HRI and, at larger radii, archival PSPC data, we find that the radial behavior of the X-ray surface brightness profile is not consistent with a simple Beta model and suggests instead three distinct components. We use a multi-component bidimensional model to study in detail these three components that we identify respectively with the cooling flow region, the galactic and the cluster halo. 
From these data we derive a binding mass distribution in agreement with that suggested by optical dynamical indicators, with an inner core dominated by luminous matter and an extended dark halo differently distributed on galactic and cluster scales. The HRI data and a preliminary analysis of {\it Chandra} public data, allow us to detect significant density fluctuations in the halo.
We discuss possible non-equilibrium scenarios to explain the hot halo structure, including tidal interactions with neighboring galaxies, ram stripping from the intra-cluster medium and merging events.  

In the innermost region of NGC 1399, the comparison between the X-ray and radio emission suggests that the radio emitting plasma is displacing and producing shocks in the hot X-ray emitting gas. We do not detect the nuclear source in X-rays and we pose an upper limit of $\sim 4\times 10^{39}$ erg s$^{-1}$ (0.1-2.4 keV) to its X-ray luminosity.

We found that the NGC 1404 halo is well represented by a single symmetric Beta model and follows the stellar light profile within the inner 8 kpc. The mass distribution is similar to the `central' component of the NGC 1399 halo. At larger radii ram pressure stripping from the intra-cluster medium produces strong asymmetries in the gas distribution. 

Finally we discuss the properties of the point source population finding evidence of correlation between the source excess and NGC 1399.

\end{abstract}
\keywords{X-rays: individual (NGC 1399, NGC 1404)---galaxies: halos---galaxies: clusters: individual (Fornax)---galaxies: jets---radio continuum: galaxies}

\section{INTRODUCTION}
NGC 1399 is the central dominant galaxy of the Fornax cluster. Due to its proximity (19 Mpc for $H_0=75$ km s$^{-1}$ Mpc$^{-1}$), this very regular, almost spherical (E0; \citealt{Ferg89}) cD galaxy has been extensively studied in a wide range of wavelengths, ranging form radio to X-rays. The optical radial profile, 
first studied by \cite{schom86} and later in more detail by \cite{kill88}, reveals a large halo extending up to 250 kpc. The galaxy is surrounded by a large number of globular clusters, 10 times in excess with respect to those of its nearer companion NGC 1404 and of the Fornax galaxy NGC 1380 \citep{Kiss99}. Dynamical studies of the stellar population of NGC 1399 indicate that the central regions of the galaxy are characterized by a high velocity dispersion, decreasing with radius, and slow rotation, as expected in giant ellipticals \citep{gra98}. At larger radii the velocity dispersion starts increasing again, as shown by studies of the planetary nebulae \citep{Arn94} and globular clusters dynamical properties \citep{Grill94}. This suggests a different dynamical structure, than that of the inner stellar body, for the galaxy envelope and the globular cluster population. 

NGC 1399 hosts a weak nuclear radio source \citep{kbe88} with radio luminosity of $\sim 10^{39}$ ergs s$^{-1}$ between $10^7$ and $10^{10}$ Hz. The radio source has two  jets ending in diffuse lobes, confined in projection within the optical galaxy. \cite{kbe88} suggest that the radio source is intrinsically small and confined by the thermal pressure of the hot ISM. 

X-ray data have shown the presence of an extended hot gaseous halo surrounding NGC 1399. Observations made with the {\it Einstein} IPC by \cite{Kim92} constrained the gas temperature to be kT$>$1.1 keV. The {\it Einstein} IPC data were used also by \cite{kill88} to derive the mass distribution of the galaxy. They found that both models with and without dark matter were compatible with those data, depending on the assumed temperature profile.
\cite{White92} suggested the presence of a cooling flow in the center of the galaxy, depositing 0.8$\pm 0.6$ M$_{\odot}$ y$^{-1}$.
Ginga observations \citep{ikebe92} led to the detection of extended emission out to a galactocentric radius of $\sim$250 kpc. \cite{ser93} constrained the X-ray temperature to be 1.0$<$kT$<$1.5 keV with the Broad Band X-ray Telescope. 
\citet[ hereafter RFFJ]{rang95}, using the Rosat PSPC, have studied in detail the temperature profile of the hot inter-stellar medium (ISM) out to 220 kpc finding an isothermal profile 
(0.9$<$kT$<$1.1) from 7 to 220 kpc, and a central cooling flow (0.6$<$kT$<$0.85) of at least 2 M$_{\odot}$ y$^{-1}$.

\cite{ikebe96} were able to identify with ASCA the presence of different components in the X-ray halo, associated respectively with the galaxy and the Fornax cluster potential.
\cite{jones97} used the better resolution of the Rosat PSPC for studying in detail the cluster X-ray structure and finding a total binding mass of (4.3-8.1)$\times 10^{12}$ M$_{\odot}$ within 100 kpc, with a mass-to-light ratio increasing from 33 M$_{\odot}$/L$_{\odot}$ at 14 kpc to 70 M$_{\odot}$/L$_{\odot}$ at 85 kpc. They also found a metal abundance of 0.6 solar.

\cite{Buo99} analized ASCA data finding that either a two temperature spectral model or a Cooling-Flow model with Fe solar abundances, are required to fit the thermal X-ray emission, and additional absorption is needed in the galaxy center. \cite{Matsu00} found similar near solar metallicities but argued that the metallicity inferred by ASCA spectral fits is dependent on the assumed atomic physics model and is not solved by muti-components spectral models. 

More recently \cite{Sulk01} used ROSAT data to pose an upper limit on the nuclear source brightness. The {\it Chandra} data were used by \cite{Loew01} to further constrain the nuclear source and by \cite{Ang01} to study the point source population hosted by globular clusters.

In this paper we present the results of a deep image of the NGC 1399--NGC 1404 field obtained from data collected between 1993 and 1996 with the ROSAT High Resolution Imager (HRI, for a description see \citealp{Dav96}). We take advantage of the $\sim 5"$ resolution of the HRI to study in detail the structure of the galactic halo and relate the results to those obtained at larger scales with poorer resolution instruments. We study the interactions between the nuclear radio source and the galactic halo and discuss the properties of the discrete sources population. A preliminary analysis of {\it Chandra} data supports the Rosat results.\\
Throughout this paper we adopt $H_0=75$ km s$^{-1}$ Mpc$^{-1}$ and a distance of 19 Mpc (1'=5.5 kpc).

\section{OBSERVATIONS AND DATA ANALYSIS}
The NGC 1399 field, including NGC 1404, was observed at three separate times
with the ROSAT HRI: in February 1993, 
between January and February 1996 and between July and August of the same year.
The total exposure time is 167.6 Ks (Table \ref{observations}).
The data were processed with the  SASS7\_8 and SASS7\_9 versions of the ROSAT standard analysis software (SASS).
For our data analysis, we used the IRAF/XRAY and CIAO packages developed
at the Smithsonian Astrophysical Observatory and at the {\it Chandra} X-ray Center (CXC), and other specific software as mentioned in the text.
%################################################################
\begin{table*}[t]
\begin{center}
\footnotesize
\caption{Rosat HRI observations of NGC 1399.\label{observations}}
\begin{tabular}{ccccrcc}
\\
\tableline
\tableline									
name & \multicolumn{2}{c}{Field center} & sequence id. &
 Exp.time & obs. date & P.I.\\
& R.A. & Dec & & (sec) & &\\
\tableline
NGC 1399  & 03$^{\rm h}$38$^{\rm m}$31$^{\rm s}$ & --35$^\circ$27'00'' & RH600256n00 & 7265 & 1993 Feb 17 & D.-W. Kim\\
''  & '' & '' & RH600831n00 & 72720 & 1996 Jan 04-1996 Feb 23 & G. Fabbiano\\
''  & '' & '' & RH600831a01 & 87582 & 1996 Jul 07-1996 Aug 26 & G. Fabbiano\\
\tableline
\end{tabular}

\end{center}
\end{table*}
%################################################################ 

\subsection{Aspect Correction}
\label{Aspect corrections}
\begin{figure}[t]
\centerline{\psfig{figure=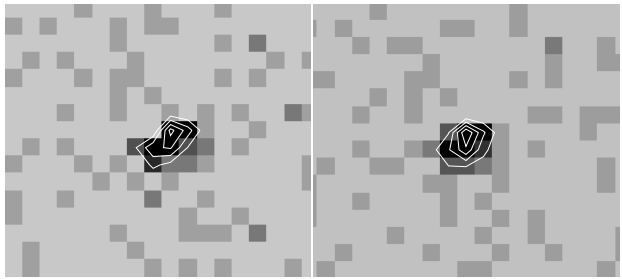,angle=180,width=0.5\textwidth}}
\caption{Comparison between the aspect of the brightest pointlike source in the NGC 1399/1404 field (No.8 in Table \ref{source_tab}) before (left) and after (right) correction. The pixel size is 4". Contour levels are spaced by $5.9\times 10^{-2}$ cnts arcmin$^{-2}$ s$^{-1}$ with the lower one at $6\times 10^{-2}$ cnts arcmin$^{-2}$ s$^{-1}$.}
\label{asp_corr}
\end{figure}

The HRI data processed with the SASS versions prior to SASS7\_B of March 1999 (as it happens for our data) suffer from an error in the aspect time behavior  \citep{Har99}. 
This translates into an error on the position of the incoming photons and thus it results in  a
degradation of the Point Response Function (PRF). Because high resolution analysis is the primary objective of this work we run the correction routine ASPTIME (F. Primini 2000, private communication) on the data, to improve the aspect solution.
Visual inspection of the brightest pointlike sources in our field (Figure \ref{asp_corr}) demonstrates the improvement in the image quality.

%#############################################
%#############################################

Additional problems that may preclude the attainment
of the HRI potential resolution are the imperfect correction of the spacecraft wobble and the wrong tracking of
reference stars due to the variable pixel sensitivity across the detector \citep{Harris98}.
We followed the procedure suggested by Harris and collaborators, of dividing the observations in time bins (OBI) and realigning these segments by using the centroids of a reference pointlike source in each OBI.

Our data can be divided in 4 OBIs in RH600256n00, 43 in RH600831n00 and 33 in RH600831a01.
We checked each OBI individually using the brightest point source in the field 
(No.8 in Table \ref{source_tab}). We found that in the first observation (RH600256n00) the OBI are well aligned and no correction is needed.
The second observation (RH600831n00) has 7 OBIs out of 43 that are 7"-10" displaced; however, the overall PRF is 6" FWHM so that just a slight improvement can be obtained dewobbling the image. The third observation (RH600831a01) has a 8" PRF but the signal to noise ratio of the individual OBIs is worse than in the previous observations so that the dewobbling procedure cannot be applied.
Given these results, we decided not to apply the dewobbling correction, and to
take as good a resulting PRF, in the composite image, of $\simeq$7" FWHM.

\subsection{Composite Observation and Exposure Corrections}
\label{Composite observation}
\begin{figure*}[t]
\centerline{\psfig{figure=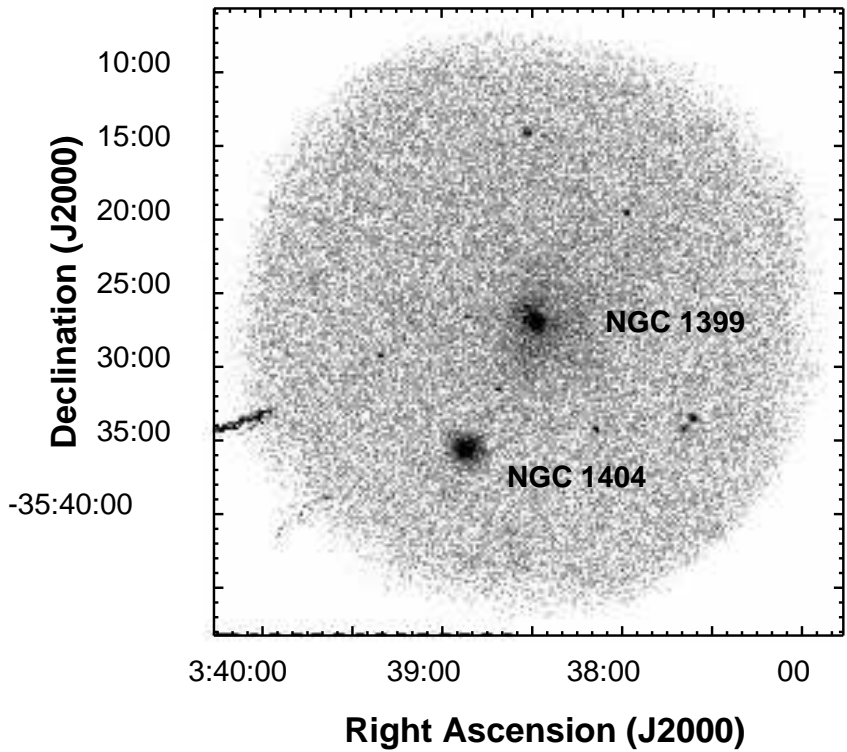,angle=0,width=.8\textwidth}}
\caption{The `raw' NGC 1399/NGC 1404 HRI field. The composite image is here displayed after it was rebinned to 5 arcsec/pixel.
Even without exposure correction the extended emission surrounding NGC 1399 is clearly visible.
The elongated features in the lower left
corner are due to the presence of ``hot spots'' on the detector.}
\label{NGC1399comp}
\end{figure*}

The aspect-corrected observations were co-added to obtain a ``composite
observation''. To make sure that pointing uncertainties did not degrade the image we used the centroids of three bright pointlike sources in the field
(No.8, 19 an 24 in Table \ref{source_tab}) to align the images. 
The applied corrections were all within 2 arcsec. 
The resulting composite image is shown in Figure \ref{NGC1399comp}.
%#############################################
%#############################################
This image was then corrected for vignetting and variations of
exposure time and quantum efficiency across the detector  by
producing an ``exposure map'' for each observation with the software
developed by S.L. Snowden \citep[ hereafter SMB]{Snow94}. 

\subsection{Brightness Distribution and X-ray/Optical Comparison}
\label{brightness}
To study the large-scale brightness distribution of the NGC 1399 field we rebinned the exposure corrected data in 5''$\times$5'' pixels. We then adaptively smoothed the image with the CXC CIAO {\em csmooth} algorithm which convolves the data with a gaussian of variable width (depending on the local signal to noise range of the image) so to enhance both small and large scale structures.
%#############################################
\begin{figure*}[t!]
\subfigure[]{\psfig{figure=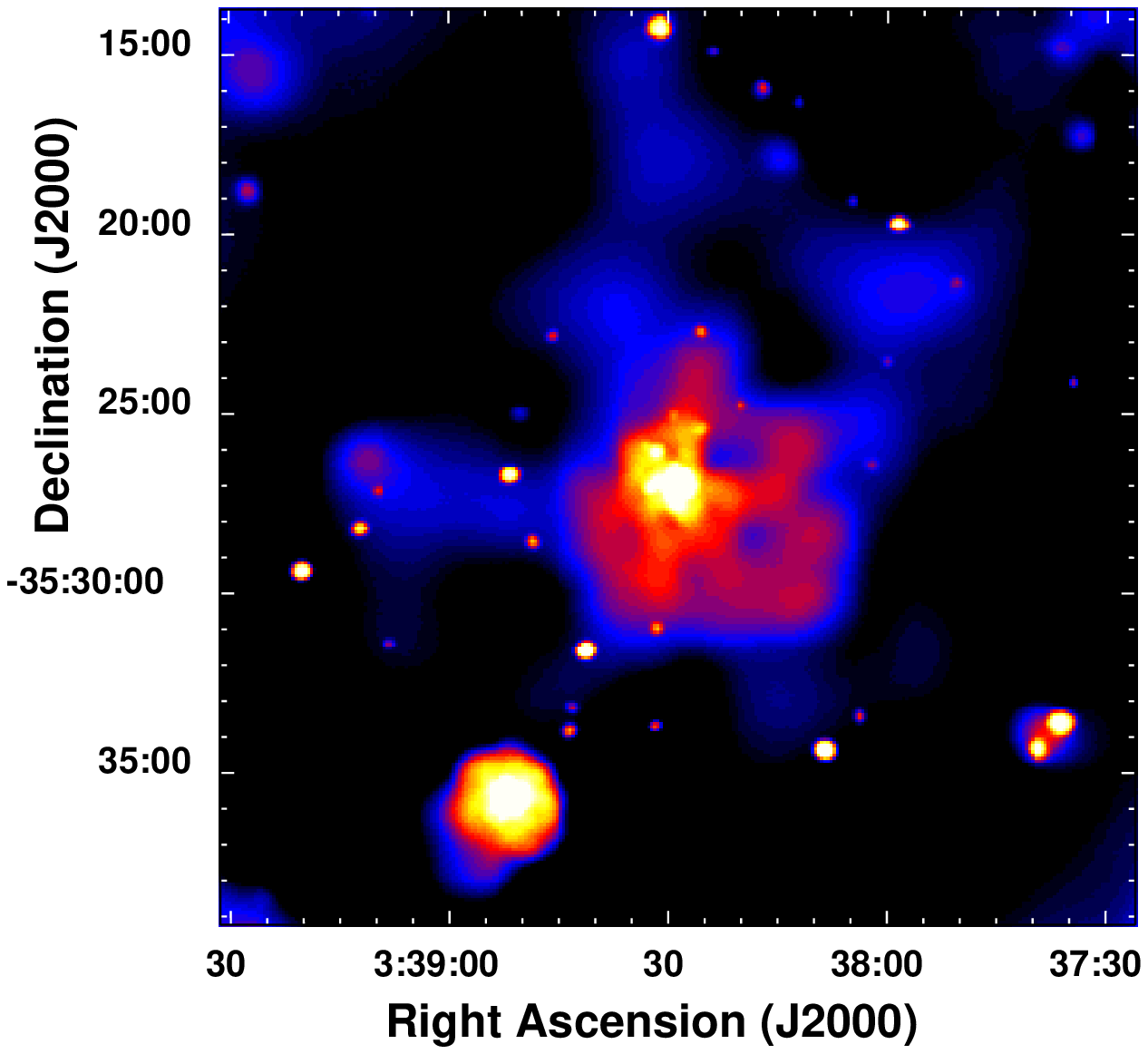,angle=0,width=0.45\textwidth}}
\subfigure[]{\psfig{figure=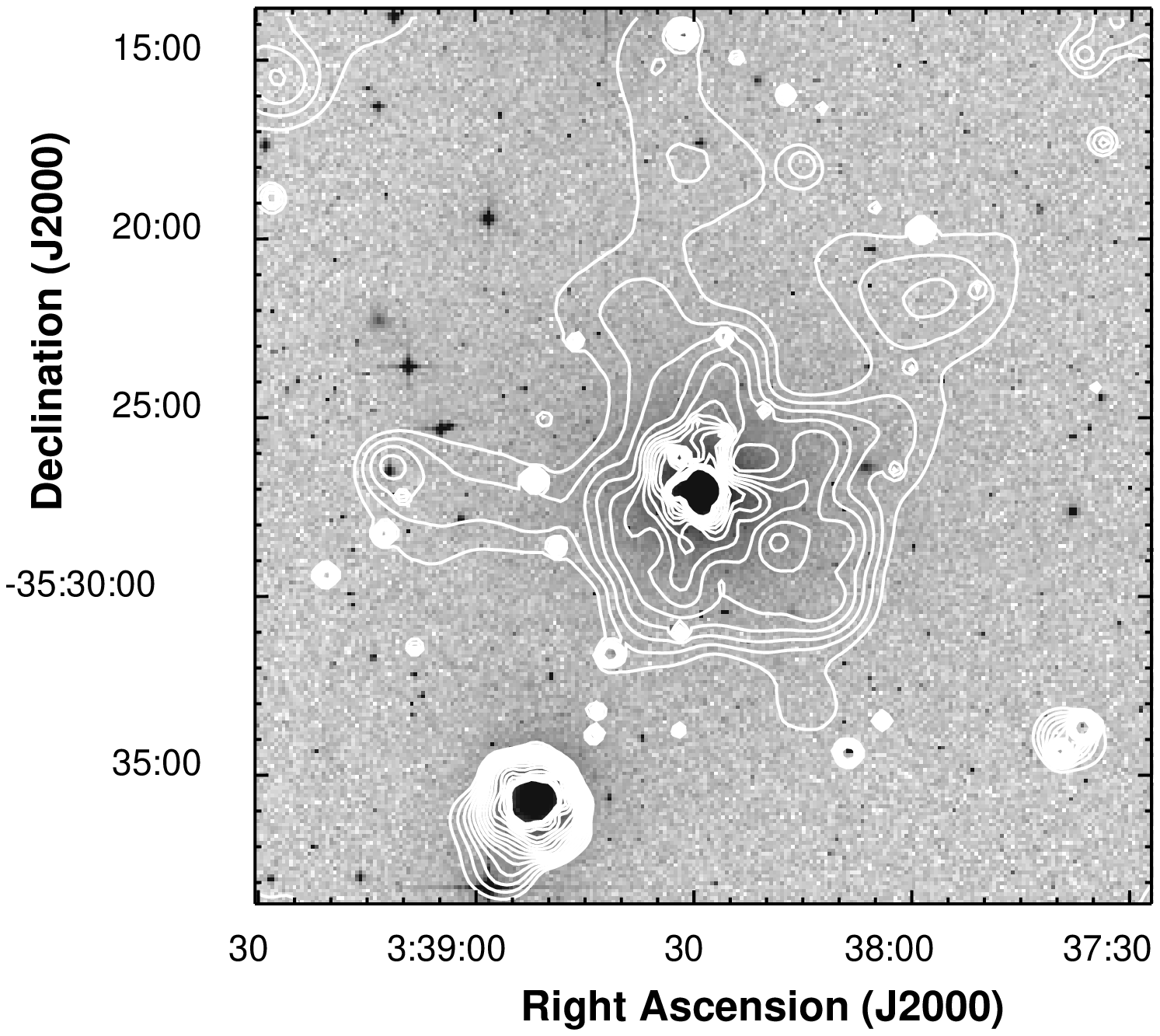,angle=0,width=0.48\textwidth}}
\caption{(a) $5\times 5$ arcsec/pixel adaptively smoothed image of the central part of the NGC 1399/1404 HRI field. Colors from black to yellow represent logarithmic X-ray intensities from $5.6\times 10^{-3}$ to $3.58\times 10^{-1}$ cnts arcmin$^{-2}$ s$^{-1}$.(b) X-ray brightness contours overlaid on the 1 arcsec/pixel 
DSS image (logarithmic grayscale). Contours are spaced by a factor 1.1 with the lowest one at $6.1\times 10^{-3}$ cnts arcmin$^{-2}$ s$^{-1}$. The X-ray emission peak is centered on the optical galaxy for both NGC 1399 and NGC 1404. In the case of NGC 1404 the X-ray isophotes are consistent with the optical distribution while in NGC 1399 the X-ray emission extends further out than the optical one.}
\label{csmooth}
\end{figure*}
%#############################################

The resulting image (Figure \ref{csmooth}a) shows a complex X-ray morphology.
The center of the image is occupied by the extended halo of NGC 1399. 
The galaxy possess a central emission peak and an external extended and asymmetric halo. This halo is not azymuthally symmetric with respect to the X-ray peak, but it extends more on the SW side. The X-ray surface brightness distribution of the halo appears filamentary, with elongated structures and  voids.
As it can be seen from a comparison with the optical Digitized Sky Survey (DSS) image (Figure \ref{csmooth}b), while the X-ray emission peak is centered on the optical galaxy, the X-ray halo extends radially much further than the optical distribution. Moreover most of the features seen in the X-ray image have no direct optical counterpart.

The X-ray emission of NGC 1404 instead is almost symmetric and largely consistent with the optical distribution, with the exception of the SE tail reported by \cite{jones97} in the PSPC data, that may be the signature of ram pressure stripping of the NGC 1404 corona infalling toward the denser NGC 1399 cluster halo.

A number of pointlike source in the field is also evident in Figure \ref{csmooth}a. The analysis of these sources is discussed in $\S$ \ref{sources}.

%#############################################
\begin{figure}[t]%[]
\centerline{\psfig{figure=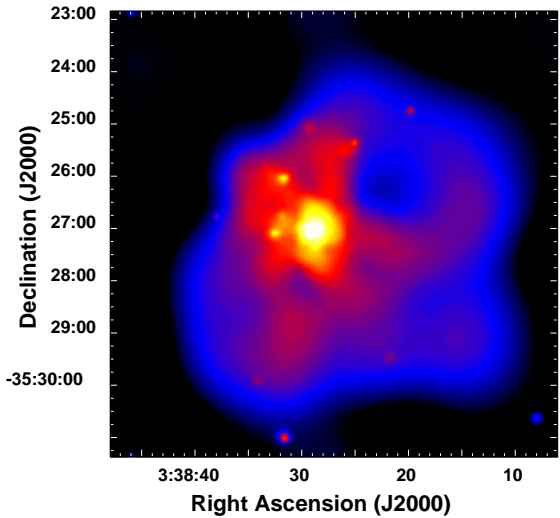,angle=0,width=0.5\textwidth}}
\caption{Adaptively smoothed image of the inner NGC 1399 halo. Colors from black to yellow represent logarithmic X-ray intensities from $6.4\times 10^{-3}$ to $7.7\times 10^{-1}$ cnts arcmin$^{-2}$ s$^{-1}$. }
\label{centbox}
\end{figure}
%#############################################

We examined the inner halo of NGC 1399 in greater detail using a 1"/pixel resolution. The adaptively smoothed image (Figure \ref{centbox}) reveals an elongation of the inner halo structure in the N-S direction plus a large arc protruding to the West side of the halo. Several voids are present in the X-ray distribution, the largest being 1.5 arcmin NW of the emission peak. The central peak appears circular with at most a slight N-S elongation.

These features are all above $3\sigma$ significance, since $3\sigma$ was the minimum significance level requested for an intensity fluctuations to be smoothed on a given scale. Some features, as the SW clump centered at RA, Dec=$3^{\rm h}38^{\rm m}15^{\rm s}$,--35$^\circ$29'00'' or the one at $3^{\rm h}39^{\rm m}10^{\rm s}$,--35$^\circ$26'30'', are also detected with a wavelets detection algorithm (see $\S$ \ref{sources}). We can rule out that features on scales of 1 arcmin are of statistical nature, because their significance can be as high as $6\sigma$ (see $\S$ \ref{model}). These brightness fluctuations must be due to the local physical conditions of the hot gaseous halo.

\subsection{Radial Brightness Profiles}

\subsubsection{NGC 1399}
\label{NGC 1399}
As a first step in the quantitative study of the X-ray emission, we created a radial profile from the HRI data, assuming circular symmetry. Count rates from the exposure-corrected composite image, were extracted in circular annuli centered on the X-ray centroid RA, Dec=$3^{\rm h}38^{\rm m}28^{\rm s}.9$, --35$^\circ$27''02'.1. We took care to remove the contribution of all the detected point-like sources by excluding circles within the $3\sigma$ radius measured by the wavelets algorithm ($\S$ \ref{sources}), from the source centroid. We also excluded a 150" circle centered on NGC 1404 (see $\S$ \ref{NGC 1404}).

\begin{figure}[t]
\centerline{\psfig{figure=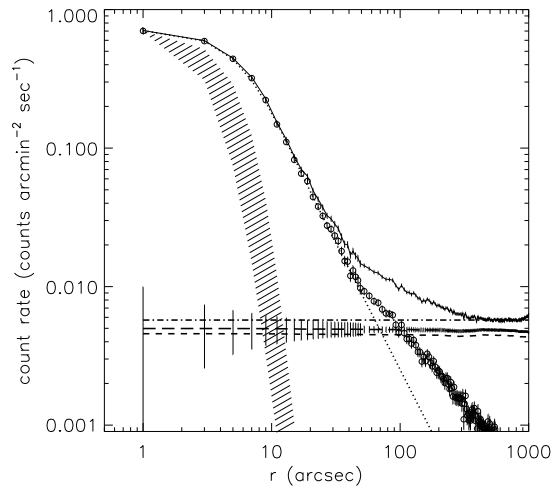,angle=0,width=0.45\textwidth}}
\caption{HRI radial profile of NGC 1399 X-ray surface brightness (continuous line) after excluding all point sources ($\S$ \ref{sources}) and NGC 1404. The dot-dashed line represents the flattening level of the HRI profile measured in the 500''-850'' annulus.
Short and long dashed lines represent respectively the SMB background level before and after rescaling to match PSPC counts.
The rescaled background was used to derive the background-subtracted counts (open circles) and the best-fit Beta model within 50" shown in Figure \ref{fit} (dotted line). 
The HRI PRF range is represented by the shaded region. Radial bins are 2" wide up to r=50", and 5" wide at larger radii.}
\label{profile}
\end{figure}

The X-ray brightness profile is shown as a continuous line in Figure \ref{profile}. The emission extends out to $\sim$ 500'' (46 kpc at the assumed distance of 19 Mpc). The radial profile flattens out past 500" (dash-dotted line in Figure \ref{profile}) suggesting that we have reached the field background level and that any residual halo emission is below our sensitivity limit. Nevertheless we know from previous investigations of the Fornax cluster \citep{jones97,ikebe92,kill88} that extended X-ray emission, bright enough to be detected in our data, is present at galactocentric radii $>500''$. We thus expected to see a smooth gradient in the X-ray profile continuing past 500''.
We checked to see if a very high background was present in some OBIs
or either in some of the HRI spectral channels but we found none that 
could explain our lack of sensitivity. A possible explanation is that we didn't reach the background level around 500'' and the actual background is much lower but, due to uncertainties in the exposure correction near the edge of HRI field of view, we are not able to see the expected decline in the X-ray emission.

%#############################################
%#############################################

\begin{figure}[t]
\centerline{\psfig{figure=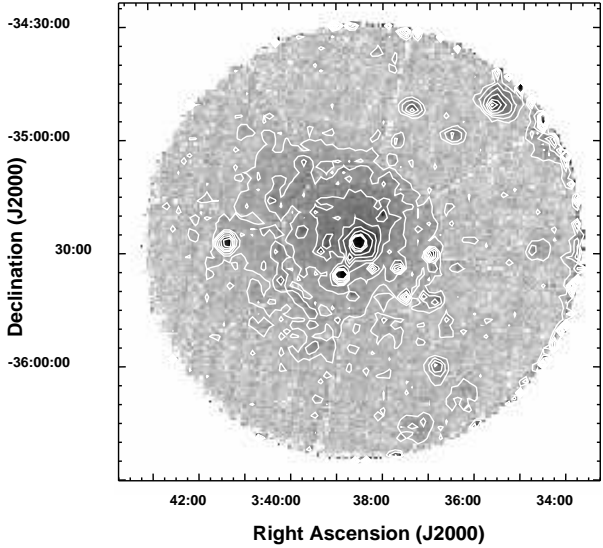,angle=0,width=0.55\textwidth}}
\caption{Exposure corrected PSPC image of the NGC 1399/1404 field. The two brightest central emission peaks represent the two galaxies (NGC 1399 and NGC 1404) visible also in the HRI field. X-ray contours, spaced by a factor of 1.3 with the lowest level at $3.5\times 10^{-3}$ cnts arcmin$^{-2}$ s$^{-1}$, clearly show the presence of an extended halo much larger than the HRI FOV (central $40\times 40$ arcmin). Radial ribs are due to the PSPC support structure.}
\label{PSPC}
\end{figure}

We used the SMB software to calculate a background map for this field. The radial profile derived from this background map is shown as a small dashed line in Figure \ref{profile}.
The SMB model however may underestimate the ``true'' HRI background somewhat because it models only the charged particle contribution. We therefore corrected this value using the 52 ks Rosat PSPC observation RP600043N00 centered on NGC 1399, taken on the 1991 August 15 (Figure \ref{PSPC}). Even if the PSPC observation is much shorter than our HRI image, the higher sensitivity and lower background of this instrument allows a better study of the large scale X-ray emission.
%#############################################
%#############################################
%#############################################
\begin{figure}[t]
\centerline{\psfig{figure=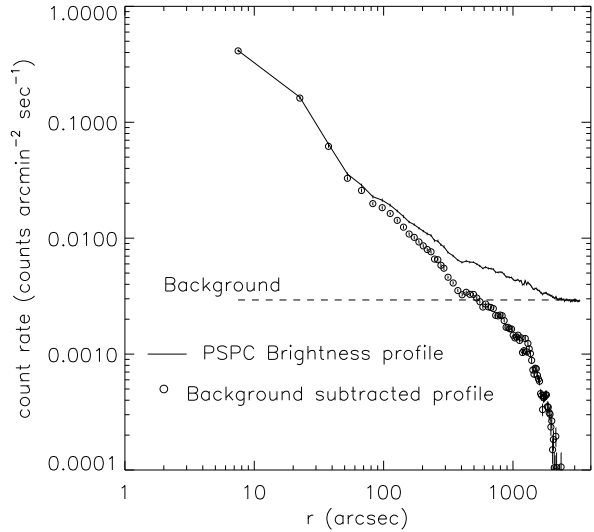,angle=0,width=0.5\textwidth}}
\caption{Exposure corrected PSPC radial profile of NGC 1399 (continuous line). 
The dashed line represents the background level measured in the 2400-3000 arcsec region, used to derive the background-subtracted counts (open circles).
The X-ray halo clearly extends at much larger radii than visible in the HRI FOV but the central region is not well resolved due to the larger PRF.}
\label{PSPC_prof}
\end{figure}
%#############################################

The exposure corrected PSPC profile is shown in Figure \ref{PSPC_prof}. The background level (dashed line) was estimated from the 2400''-3000'' annulus. 
We compared our radial profile to the one derived by \cite{jones97} and RFFJ, who made use of the same data, finding consistent results. 
We then rescaled the SMB background to find the best agreement between HRI and PSPC counts in the 50-400 arcsec region, after rebinning the HRI data to match the wider PSPC PRF. The ``true'' HRI background level found in this way (long-dashed line in Figure \ref{profile}) lies in between the flattening level of the HRI profile and the SMB background and was used to derive the HRI background-subtracted profile shown as empty circles.
This way of finding the HRI background assumes that the PSPC brightness profile is not affected by systematic errors on the background determination and can thus be used to rescale the HRI counts. However, even if the PSPC background suffered from residual errors (e.g. due to uncertainties in the exposure correction at large radii), they would not significantly affect the central region profile -- used to rescale the SMB HRI background -- because there the X-ray brightness is one order of magnitude larger than the PSPC background level.

\begin{figure}[t]%[]
\centerline{\psfig{figure=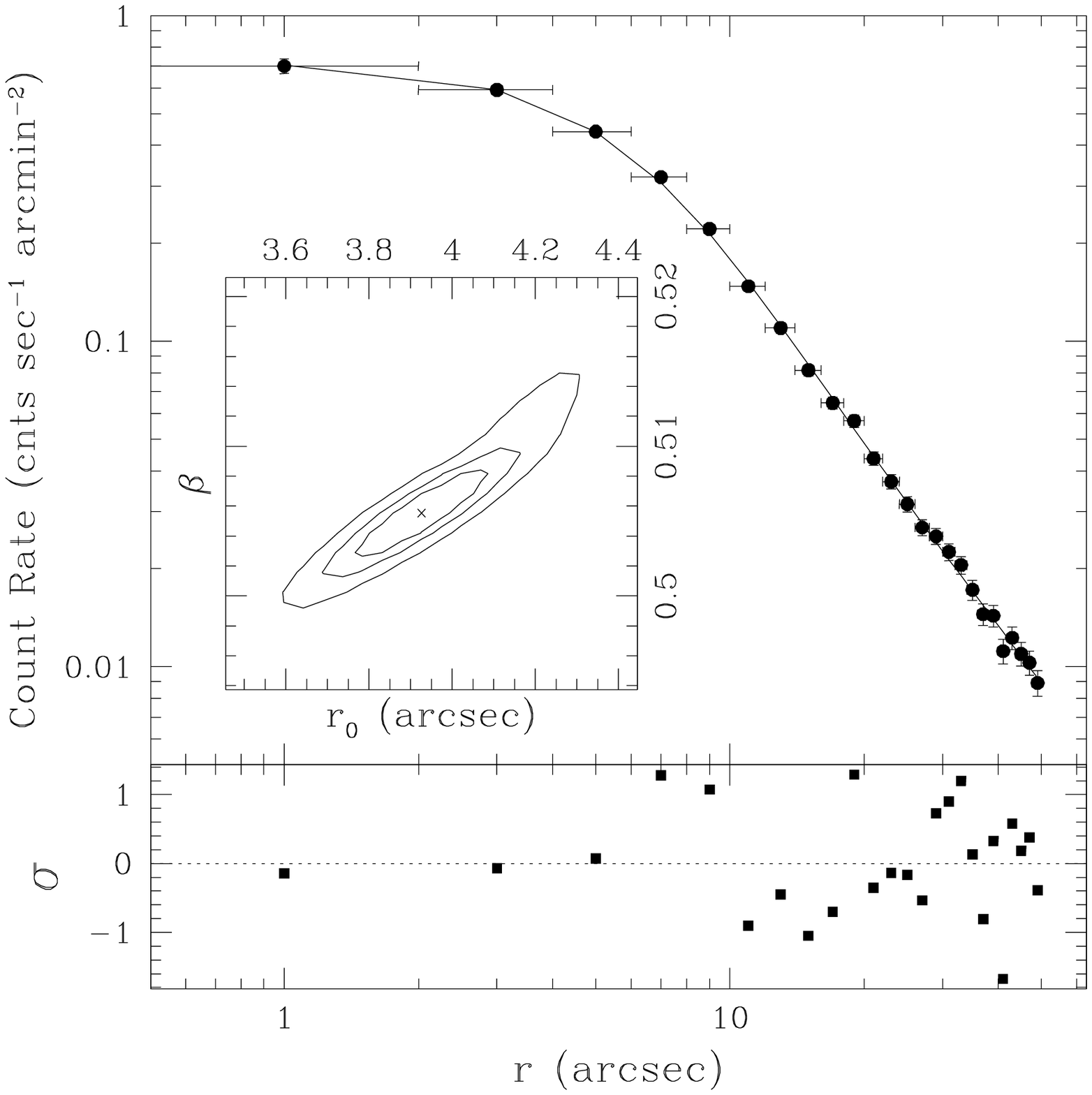,angle=0,width=0.4\textheight}}
\caption{The best-fit model and the residuals of NGC 1399 brightness profile within 50''. The 66\%, 90\% and 99\% contour levels relative to the core radius $r_0$ and slope $\beta$ are shown in the inner panel.}
\label{fit}
\end{figure}

The HRI radial surface brightness profile has a complex structure. In the `central' region, i.e. within 50'' from the emission peak, the X-ray emission is well represented by a simple Beta model:
\begin{equation}
\label{beta}
\Sigma\propto [1+(r/r_0)^2]^{-3\beta+0.5}
\end{equation}
with best fit parameters $r_0=3.93\pm 0.16$ arcsec, $\beta=0.506\pm 0.003$ and $\chi^2=14.5$ for 22 degrees of freedom (Figure \ref{fit}). To determine the best fit parameters the model was convolved with the HRI on-axis PRF \citep{Dav96}. At radii larger than 1 arcmin the surface brightness profile shows a significant excess over this model (see Figure \ref{profile}), indicating the presence of an additional `galactic' component. Restricting the fit to r $<$ 40'' to check if this more extended component contributes around 50'', gives consistent results with $r_0=3.88^{+0.20}_{-0.16}$ arcsec and $\beta=0.504^{+0.004}_{-0.003}$ ($\chi^2=11$ for 17 degrees of freedom). Thus, for $3"<r<50"$ the X-ray emission falls as $r^{-2.04\pm 0.02}$. 
As can be seen by the comparison with the HRI PRF (shaded region in Figure \ref{profile}), whose uncertainty is due to residual errors in the aspect solution \citep{Dav96}, the central component is extended and cannot be due to the presence of a nuclear point source, that in fact is not detected in our data (see $\S$ \ref{Xradio}). 
%#############################################
%#############################################
%#############################################
\begin{figure*}[t]
\centerline{
\subfigure[]{\psfig{figure=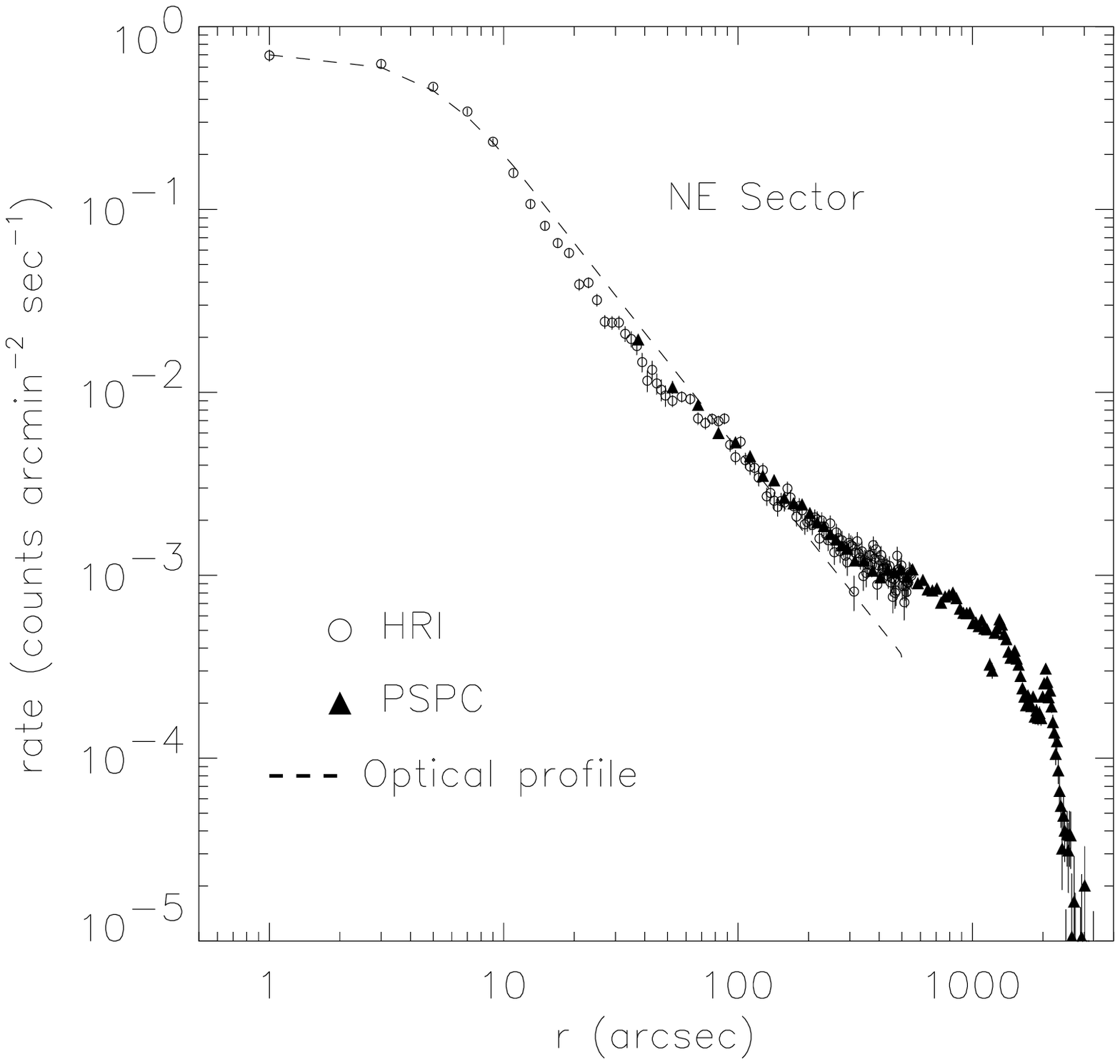,angle=0,width=0.45\textwidth}}
\subfigure[]{\psfig{figure=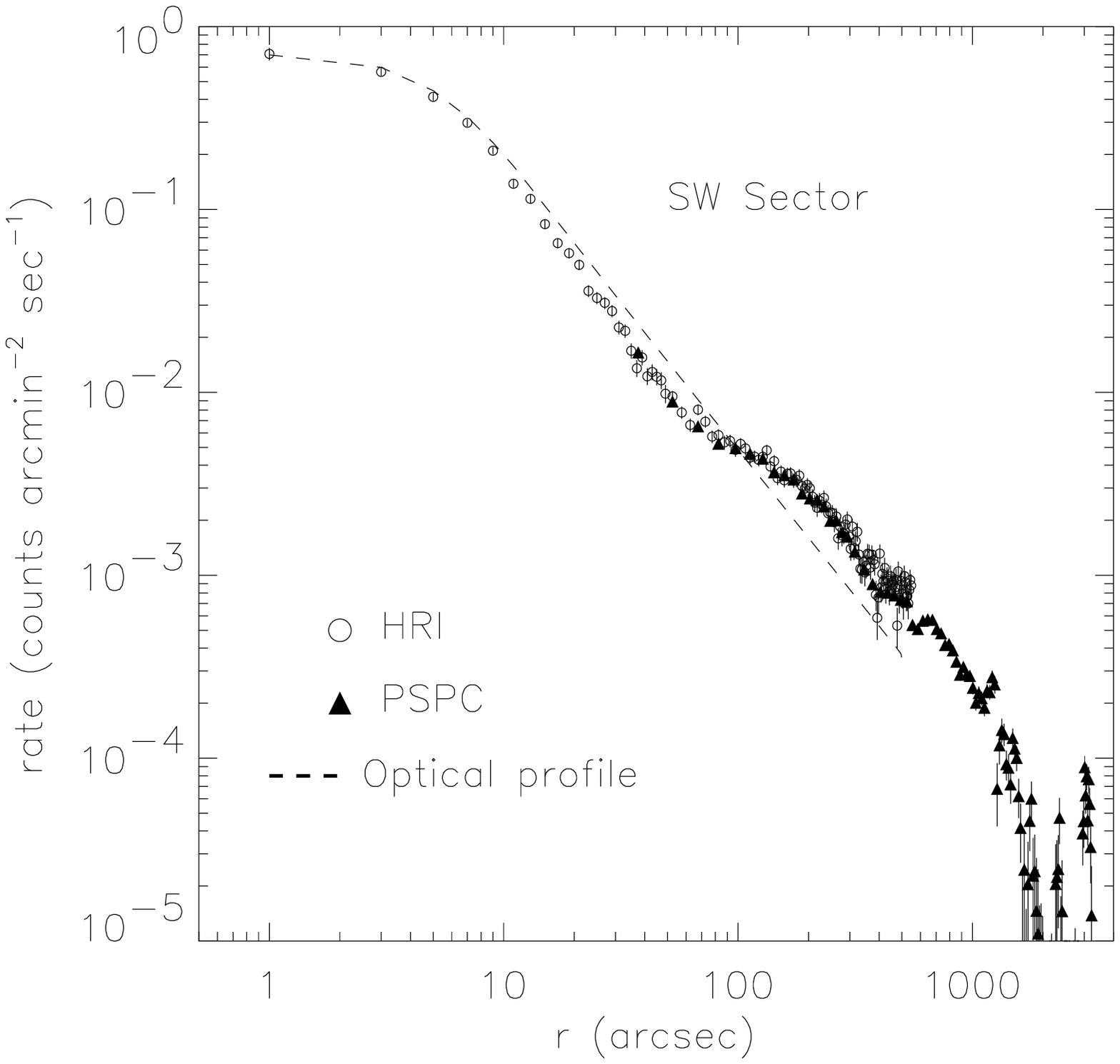,angle=0,width=0.45\textwidth}}
}
\caption{Background subtracted radial profile of NGC 1399 in the NE (a) and the SW (b) sectors. Open circles represent HRI data (2" annuli for $r<50$", 5" for $r>50"$), while filled triangles represent PSPC data (15" annuli for $r<300$", 30" for $r>300"$). PSPC counts are reduced by a factor $\sim 0.3$ to match the HRI profile; the inner 30" are not shown because of the lower resolution. The dashed line represents the optical profile (convolved with the HRI PRF) from \cite{kill88}.}
\label{pies}
\end{figure*}
%#############################################  

The PSPC profile in Figure \ref{PSPC_prof} shows that X-ray emission extends to much larger radii than visible in the HRI data and suggests the presence of a third component dominant at $r>400"$. Large azimuthal differences are seen in the X-ray profile at radii larger $\sim 1$ arcmin. We derived composite HRI-PSPC X-ray profiles in two sectors dividing the images along the line connecting the X-ray centroids of NGC 1399 and NGC 1404: a NE sector (P.A. 331.5$^{\circ}$--151.5$^{\circ}$) and a SW sector (P.A. 151.5$^{\circ}$--331.5$^{\circ}$). While the emission within 1 arcmin is azimuthally symmetric, the galactic component ($r>1'$) is much more extended in the SW direction and presents a steep decline in the NE sector, as suggested by Figures \ref{csmooth}a and \ref{centbox}. The even more extended component visible in the PSPC data follows the opposite behavior, being 3 times brighter around 1000'' on the NE side than on the SW one. While these asymmetries have been reported by \cite{jones97}, their PSPC data alone do not have enough resolution to clearly separate the central component from the external halo. Jones and collaborators used a single power-law model to fit the whole distribution, thus obtaining a value of $\sim$0.35 for the power law slope.

\subsubsection{NGC 1404}
\label{NGC 1404}
%#############################################
\begin{figure*}[t]
\begin{centering}
\subfigure[]{\psfig{figure=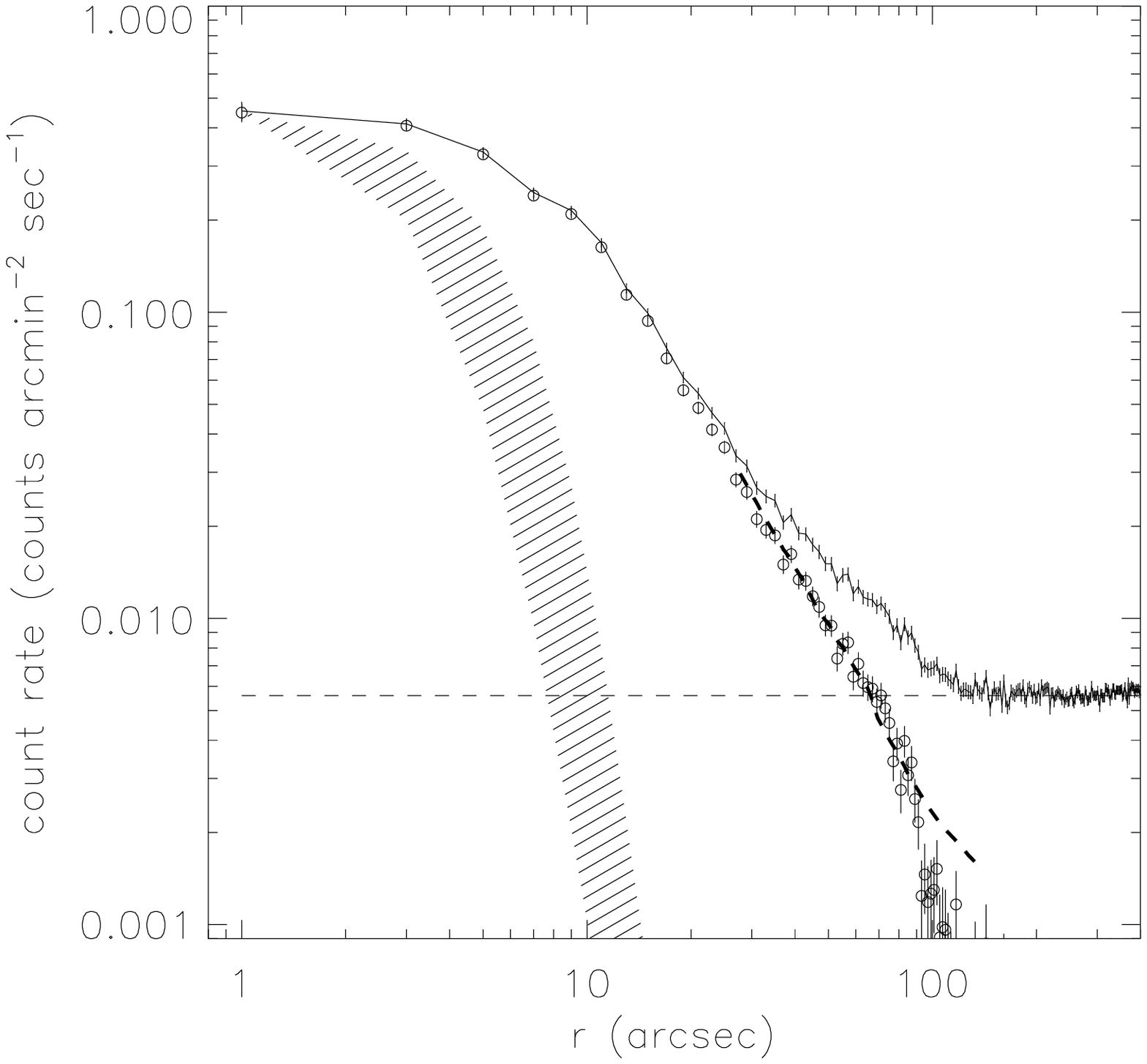,angle=0,width=0.43\textwidth}}
\subfigure[]{\psfig{figure=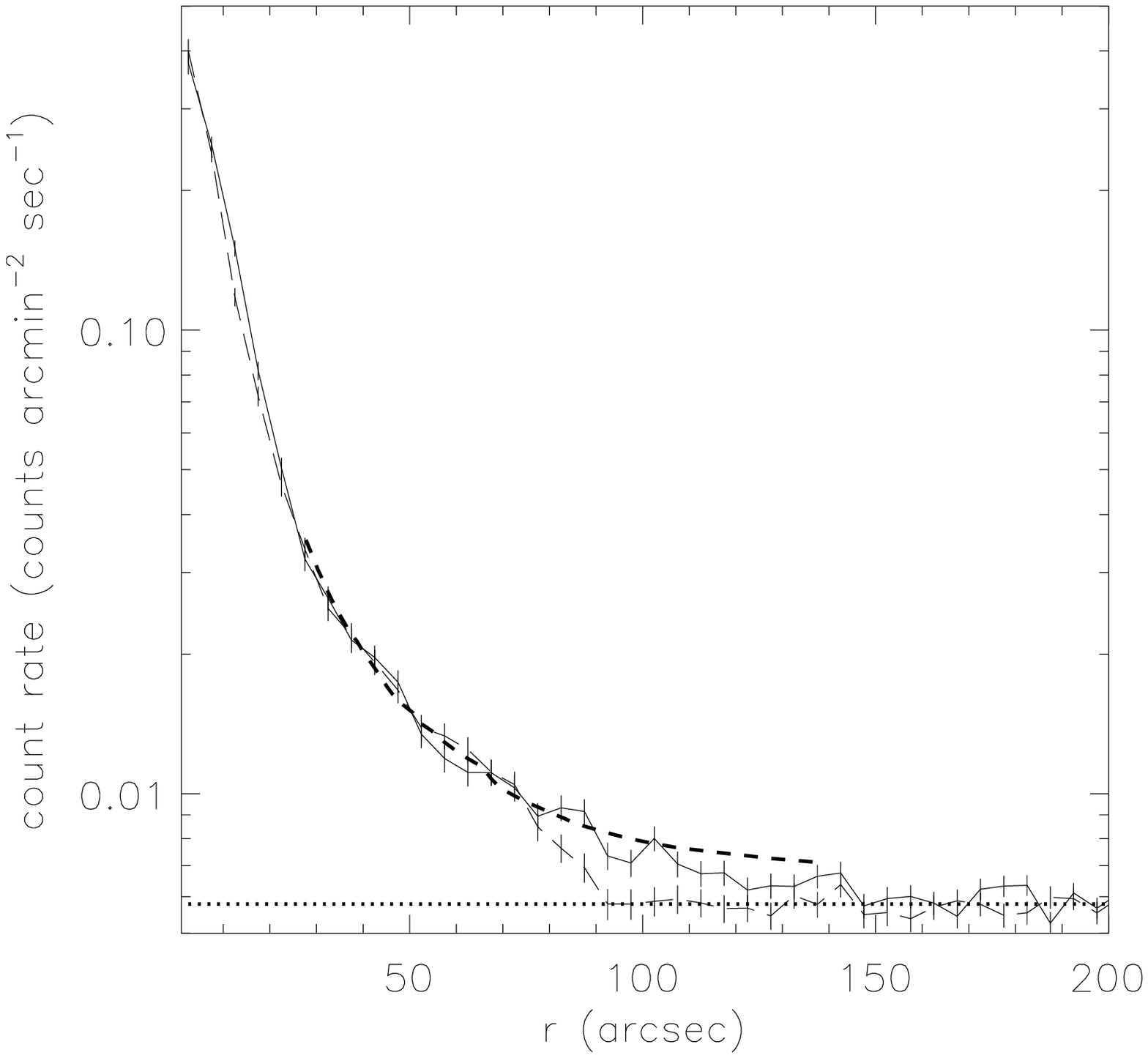,angle=0,width=0.43\textwidth}}
\caption{{\bf (a)} Radial profile of NGC 1404 extracted in 2'' annuli (continuous line). Circles represent the background subtracted profile using the count rate measured in the 200''-300'' interval (dashed line). The optical profile from \cite{Forb98} is shown as a dashed line. The HRI on-axis PRF range is represented by the shaded region.{\bf~(b)} Radial profiles extracted in 5'' annuli in the North-West (P.A. 270$^\circ$-360$^\circ$, thin dashed line) and the South-East (P.A. 90$^\circ$-180$^\circ$, continuous line) sectors. The optical profiles and background levels are represented respectively by the thick dashed line and the dotted line.}
\label{1404profile}
\end{centering}
\end{figure*}
%#############################################
We derived the brightness profile of NGC 1404 extracting count rates in 2'' annuli centered on the galaxy centroid. The background determination for this galaxy is even more difficult than for NGC 1399, due to the fact that it is embedded in the halo of the dominant galaxy. Thus we had no other choice than to estimate the background level from the region between 200'' and 300'' from the galaxy (see Figure \ref{1404profile}a), obtaining $5.60\times 10^{-3}$ counts arcmin$^{-2}$ sec$^{-1}$. The measured count rates are shown in Figure \ref{1404profile}a as a continuous line while the background subtracted profile is represented by open circles.
The NGC 1404 emission extends out to 110'' ($\sim$10 kpc) from the galaxy center and shows no sign of a multi component structure as the one detected in NGC 1399. The surface brightness is circularly symmetric (cf. Figure \ref{csmooth}a) and declines with a power law profile out to 90''. 
Past this radius the effect of ram pressure stripping becomes evident as a sharp cutoff in the radial profile. In Figure \ref{1404profile}b we can see that in the North-West sector (thin dashed line) the galactic halo suddenly disappears in the X-ray background for $r>90"$ ($r>8$ kpc) while in the South-East one (continuous line) the emission from the tail, seen in the adaptively smoothed images (Figure \ref{csmooth}), extends up to $\sim 150"$ (14 kpc). 

\begin{figure}[t]
\centerline{\psfig{figure=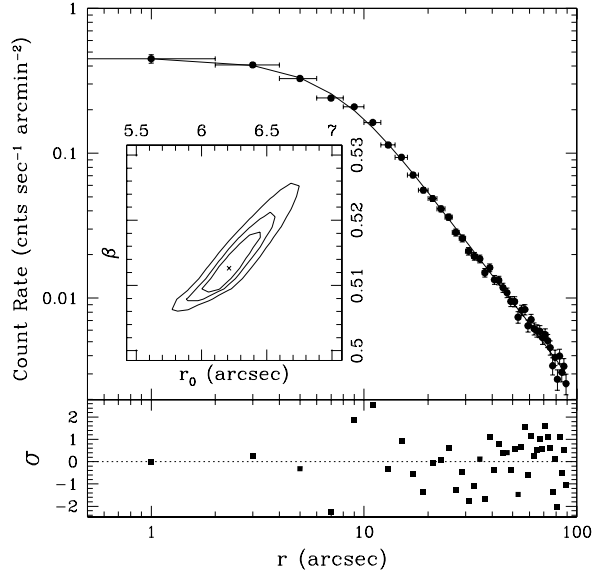,angle=0,width=0.4\textheight}}
\caption{The best-fit model and residuals of NGC 1404 brightness profile within 90''. The 66\%, 90\% and 99\% contour levels relative to the core radius $r_0$ and slope $\beta$ are shown in the inner panel.}
\label{1404fit}
\end{figure}

We fitted the observed brightness distribution, within a 90'' radius, with a Beta model. In this case the on-axis PRF we used for NGC 1399 is no longer valid because NGC 1404 is located 10 arcmin South-West of NGC 1399. Although an analytical correction of the on-axis PRF is possible \citep{Dav96}, it doesn't take into account the PRF azimuthal asymmetry.
We thus convolved the Beta model with the on-axis PRF and considered the resulting core radius as an upper limit to the correct value, obtaining
$r_0=6.21^{+0.25}_{-0.2}$ arcsec and $\beta=0.513^{+0.005}_{-0.004}$ with $\chi^2=51.5$ for 42 d.o.f. The best fit model is shown in Figure \ref{1404fit}.
%#############################################
%#############################################

\subsection{Bidimensional Halo Models}
\label{model}
Because of the complex halo structure it is not possible to obtain a satisfactory fit of the entire surface brightness distribution of NGC 1399 with a single circularly symmetric model. The {\em Sherpa} modeling and fitting application of the CIAO package allows us to adopt a more realistic approach by performing a two dimensional fit of the  X-ray images, using a complex model obtained by adding together three bidimensional Beta profiles of the form:
\begin{equation}
\label{2dmodel}
\Sigma(x,y)=\frac{A}{(1+(r/r_0)^2)^{3\beta-0.5}}
\end{equation}
where
\begin{equation} r(x,y)=\frac{x^2_{new}(1-\epsilon)^2+y^2_{new}}{1-\epsilon}\end{equation}
and
\begin{equation}x_{new}=(x-x_0)\cos(\theta)+(y-y_0)\sin(\theta)\end{equation}
\begin{equation}y_{new}=(y-y_0)\cos(\theta)-(x-x_0)\sin(\theta).\end{equation}
Expression (\ref{2dmodel}) represents the bidimensional extension of the Beta model in equation (\ref{beta}) where we also allow the model ellipticity ($\epsilon$) and position angle ($\theta$) to vary. 

%################################################################
\begin{table*}[t]%[]
\begin{center}
\footnotesize
\caption{Best-fit parameters for the bidimensional multi-component halo model.\label{fit_tab}}
\begin{tabular}{lcccccccc}
\\
\tableline									
\tableline									
Component & \multicolumn{2}{c}{Center Position} & $r_0$ &
 $\beta$ & $\epsilon$ & $\theta$ & $\chi^2_\nu$ & $\nu$ \\
& R.A. & Dec & (arcsec) & & $(1-\frac{minor~axis}{major~axis})$ & (rad) & & (d.o.f.)\\
\tableline									
Central & 03$^{\rm h}$38$^{\rm m}$28$^{\rm s}$.9 & --35$^\circ$27'01'' & 3.93\tablenotemark{a} & 0.54$\pm $0.02 & 0.0\tablenotemark{a} & 0.0\tablenotemark{a} & 1.2\tablenotemark{b} & 1612\\
Galactic & 03$^{\rm h}$38$^{\rm m}$25$^{\rm s}$.0 & --35$^\circ$27'40'' & 2125$\pm$ 825 & 41$\pm$ 25 & 0.018$\pm$ 0.018 & 1.09$^{+1.97}_{-1.09}$  & 1.2\tablenotemark{b} & 1612\\
Cluster  & 03$^{\rm h}$38$^{\rm m}$48$^{\rm s}$.2 & --35$^\circ$23'05'' & 9047$\pm$ 495 & 24$\pm $ 3 & 0.14$\pm$ 0.02 & 0.0 $\pm$ 0.1
& 1.5 & 4790\\
\tableline									
\end{tabular}
\tablenotetext{a}{The Central component parameters $r_0$, $\epsilon$ and $\theta$ have no error because they were held fixed during the fit.}
\tablenotetext{b}{The $\chi^2$ value is the same for the `Central' and `Galactic' components because they were fitted together on the HRI image.}	
\end{center}
\end{table*}
%################################################################ 
%#############################################
\begin{figure*}[t]%[]
\centerline{\psfig{figure=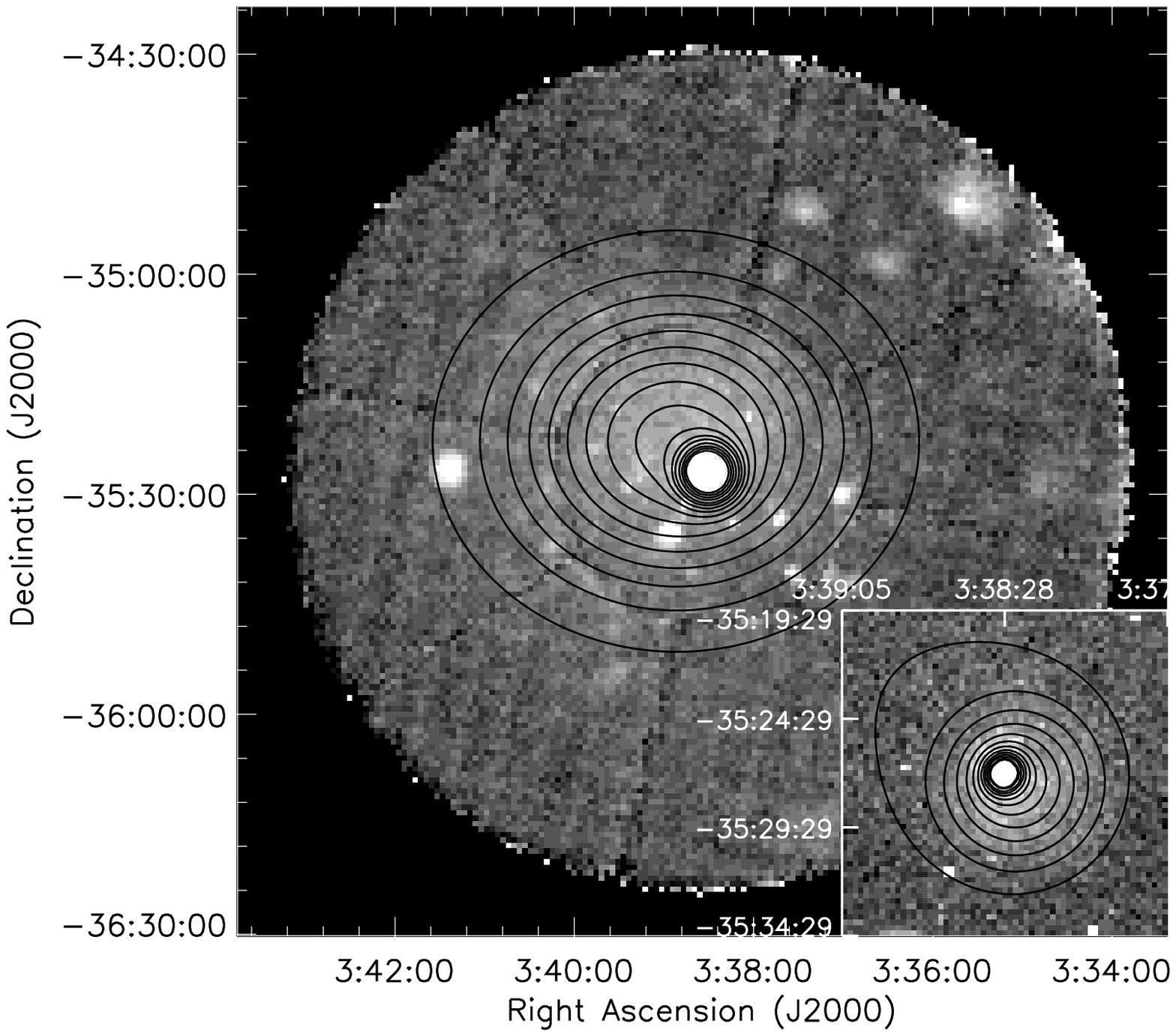,angle=0,width=0.6\textwidth}}
\caption{{\bf Main panel:} contours of the bidimensional halo model superimposed on the 40''/pixel PSPC image. The cluster component is clearly displaced with respect to the galactic halo; the central component is not visible at this resolution. Contours are spaced by a factor of 1.1 with the lowest one at 3.1$\times 10^{-3}$ cnts arcmin$^{-2}$ s$^{-1}$. 
{\bf Bottom-right panel:} bidimensional model contours overlaid on the 15''/pixel HRI image, showing in greater detail the galactic halo region. Contours are spaced by a factor of 1.1 with the lowest one at 6.0$\times 10^{-3}$ cnts arcmin$^{-2}$ s$^{-1}$.}
\label{model_map}
\end{figure*}
%#############################################

To model simultaneously the small and large scale components we used the following procedure:
{\bf (a)} we rebinned the HRI image in 15''$\times$15'' pixels, in order to have enough counts per pixel to fit the galaxy halo and to smooth out small scale asymmetries; NGC 1404 and all the detected point-like sources ($\S$ \ref{sources}) were masked out before performing the fit.
{\bf (b)} The PSPC data were rebinned in 45''$\times$45'' pixels; the regions containing the support structure where masked, together with all the sources detected in the HRI field. Because the PSPC field of view (FOV) is larger than the HRI one, to mask all the remaining PSPC sources we used the source parameters measured by the wavelets algorithm in the context of the GALPIPE project \citep[ and references therein]{mack96}, which
has produced a list of all sources present in the ROSAT PSPC archive\footnote{An online version of the GALPIPE database and documentation can be found at the D.I.A.N.A. homepage at the Palermo Observatory: http://dbms.astropa.unipa.it/}. 
{\bf (c)} We built a model composed of three bidimensional $\beta$ components represented by expression (\ref{2dmodel}) plus a constant background.
{\bf (d)} The first two components, representing the central and galactic halo emission, were fitted to the 15''$\times$15'' HRI data, fixing the third component parameters to some initial values. Because at this resolution it is not possible to properly determine all the parameters of the central component, we fixed the core radius to the best-fit value obtained from the radial fit shown in Figure \ref{profile}. While the central component shows a slight N-S elongation ($\epsilon\sim 0.05$, $\S$\ref{brightness}) at this resolution our attempts to vary also ellipticity $\epsilon$ and position angle $\theta$ resulted in no improvements of the fit, so that they were fixed to zero.
{\bf (e)} The third component, corresponding to the cluster emission, was fitted to the 45''$\times$45'' PSPC data, fixing the central and galactic halo parameters to the values obtained in step (d). The background value was fixed to the flattening level measured in $\S$\ref{NGC 1399}.
{\bf (f)} Steps (d) and (e) were repeated iteratively until the best fit parameters converged within the errors. 

The best-fit model parameters are shown in Table \ref{fit_tab}. In Figure \ref{model_map} we show the model contours superimposed on the 45''/pixel PSPC image (main panel); the nuclear region is shown in greater detail in the right-bottom panel, with contours overlaid on the 15''/pixel HRI image. 
The three components have different features: while the central one is centered on the galaxy and is highly peaked, the more extended galactic halo is displaced to the South-West thus accounting of the observed asymmetries shown in Figure \ref{pies}. In contrast with the circular symmetry of the smaller components ($\epsilon_{Central}\sim 0.05, \epsilon_{Galactic}=0.02$) the third component is elongated in the East-West direction and its center is displaced to the North-East of the optical galaxy. These results suggest a different origin of each component, that will be discussed in detail in $\S$ \ref{discussion}.

The $\chi^2$ values obtained from the fit are quite large. This is not surprising considering the large spatial anisotropy of the X-ray surface brightness (see $\S$ \ref{brightness}). The situation is worse for the PSPC data (see Figure \ref{PSPC}), also because of the poor knowledge of the exposure corrections past $\sim 1100''$ from the center of the FOV (the position of the support ring) and the large and irregular PRF at these radii that makes difficult a correct source masking.
In Table \ref{fit_tab} the `Central' and `Galactic' components have the same $\chi^2$ value because they were fitted together to the HRI data. If we
restrict the fit to the region within 1 arcmin and use higher resolution (2''/pixel) we obtain a $\chi^2_{\nu}\sim 0.6$, in agreement with the results obtained in $\S$ \ref{NGC 1399} from the profile fitting.

%#############################################
\begin{figure}[t]
\centerline{\psfig{figure=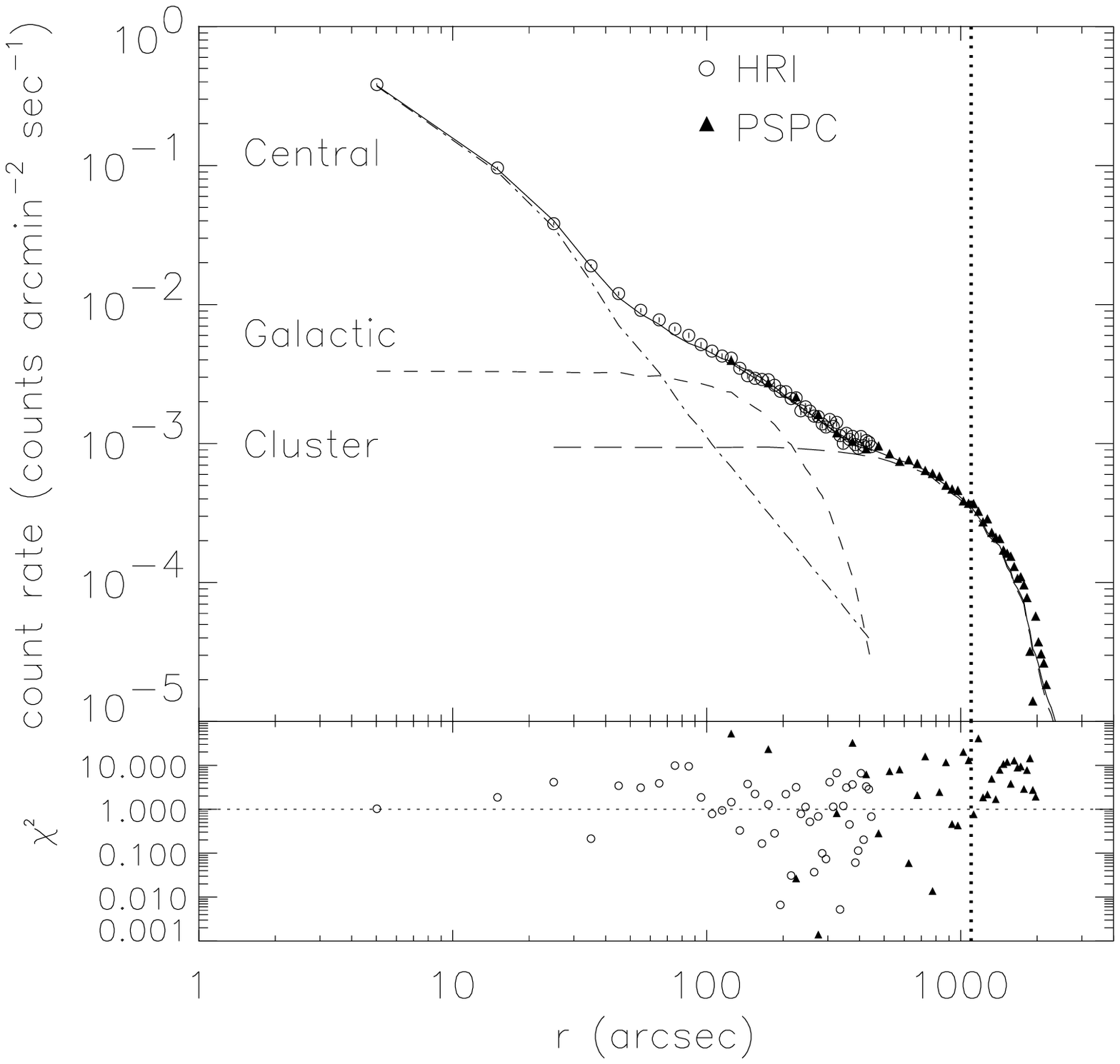,angle=0,width=0.35\textheight}}
\caption{Radial profiles compared to the bidimensional model profiles. Open circles represent HRI counts while triangles represent PSPC counts rescaled to match the HRI. The continuous line represents the best-fit model while dot-dashed, short-dashed and long-dashed lines represent respectively the central, galactic and cluster halo components. The residuals are shown in the lower panel in $\chi^2$ units. The vertical dotted line marks the position of the PSPC support ring; at larger radii the exposure correction is poorly known.}
\label{model_prof}
\end{figure}
%#############################################

The relative contribution of each component to the global profile is shown in Figure \ref{model_prof}. We must notice that here both global and individual components profiles were extracted in annuli centered on the X-ray centroid (rather than centering each one on the respective component centroid) so to allow a comparison of the observed data with the model. The bottom panel shows the radial residuals respect to the best fit model. The high $\chi^2$ values are in agreement with the values obtained from the bidimensional fit; in particular the residuals are systematically higher at radii larger than 1100'' due to the problems explained above.

%#############################################
\begin{figure}[t]
\centerline{\psfig{figure=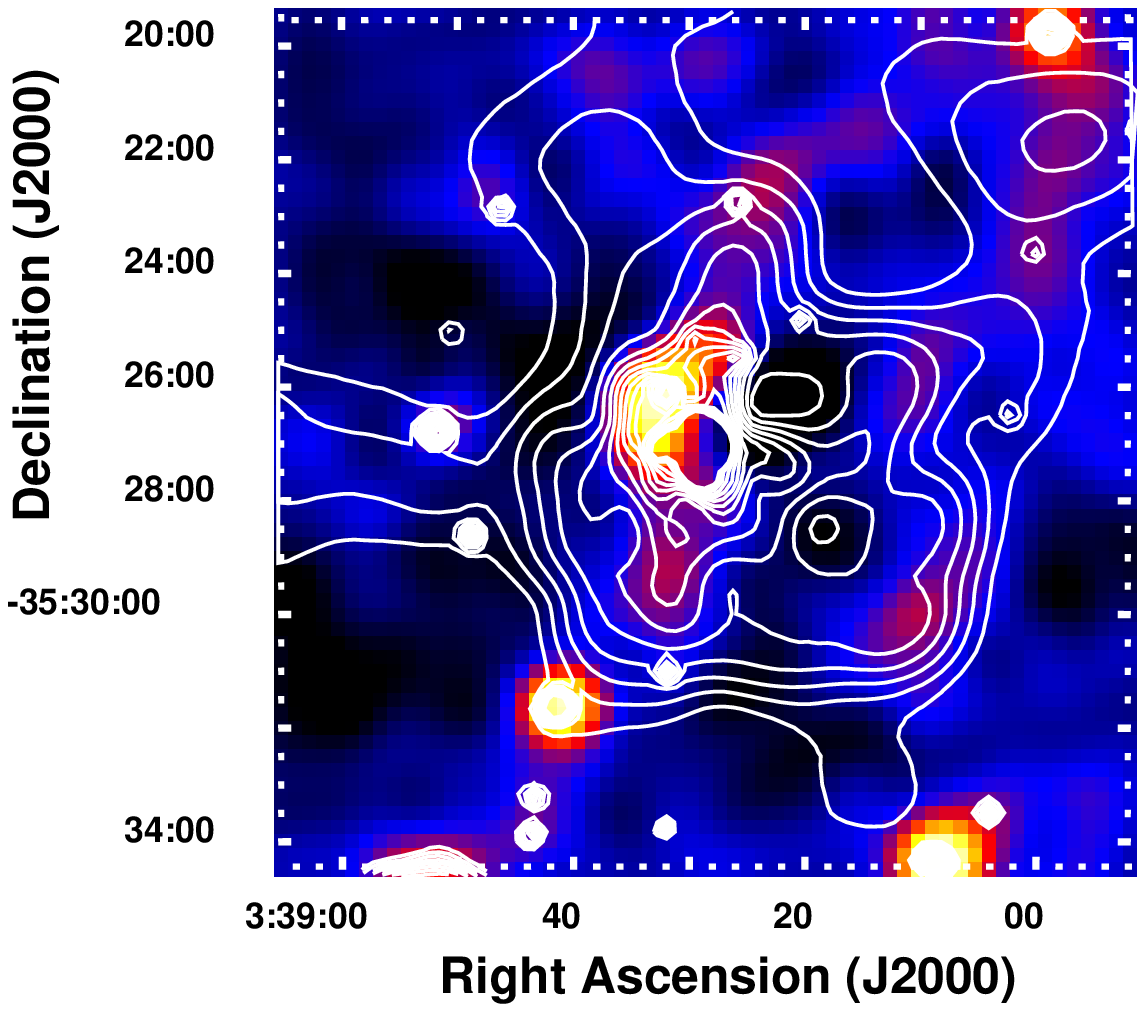,angle=0,width=0.4\textheight}}
\caption{Residuals of the bidimensional fit on the 15 arcsec/pixel HRI image, after convolving with a gaussian of $\sigma=30''$. Colors from black to white represent residuals from $-2.5 \sigma$ to $22 \sigma$. For comparison we superimposed the contour levels of the adaptively smoothed image (Figure \ref{csmooth}a).}
\label{residuals}
\end{figure}
%#############################################

The residuals of the bidimensional fit to the HRI image are shown in Figure \ref{residuals}. The presence of structures in the gaseous halo is evident and confirms the results obtained with the adaptive smoothing technique. The highest residuals are coincident with the positions of point sources detected in the field, as can be seen from the overlaid contours. Considering the fluctuations on scales of 60'' ($\sim 5$ kpc)
the statistical errors are reduced of a factor 4 with respect to the grayscale levels shown in Figure \ref{residuals}, and the significance of the residual emission is increased of the same factor. This means, for instance, that the deep `hole' on the North-West side of the halo is approximately significant at the $6\sigma$ level above the fluctuations expected from the multicomponent model, while the excess emission regions on the North and South sides of the galactic center are significant at the $3\sigma$ level. Even though the multi-component bidimensional model may not be a proper physical representation of the hot X-ray halo this result shows that the filamentary structures seen in the halo are not due to statistical fluctuation over a smooth surface brightness distribution. This result is further confirmed by the {\it Chandra} data presented in $\S$ \ref{Chandra}. 

\subsection{Density, Cooling Time and Mass Profiles}
\subsubsection{NGC 1399}
\label{dens_par}
%#############################################
\begin{figure}[th]
\centerline{\psfig{figure=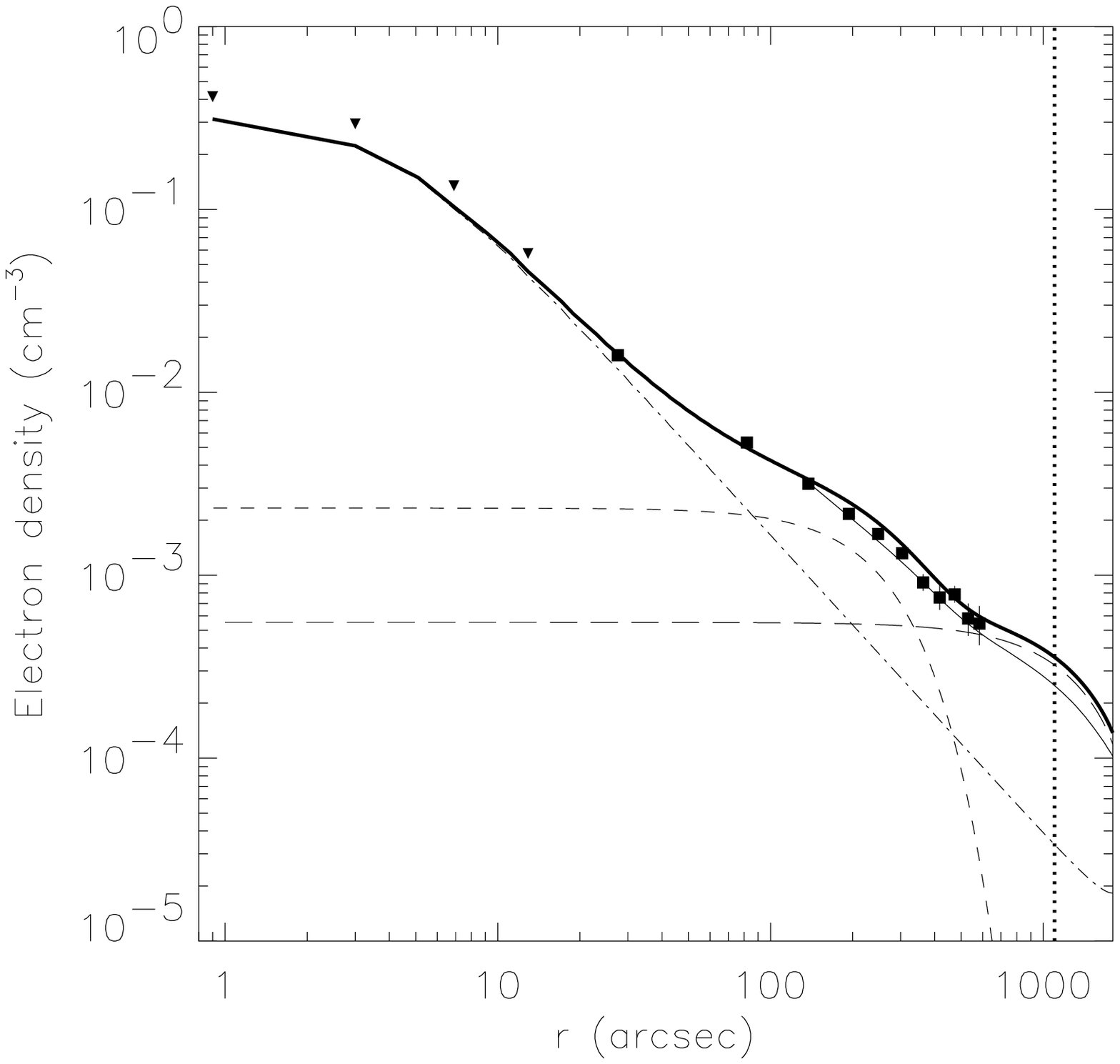,angle=0,width=0.35\textheight}}
\caption{Density profiles derived from deprojection. Dot-dashed, short dashed and long dashed lines represent respectively the central, galactic and cluster halo component. Downward triangles show the upper limit on the central component profile adopting a central absorbing column of $2\times 10^{21}$ cm$^{-2}$. The thick (thin) continuous lines show the total density ignoring (including) the center offset of the different components (see discussion in text). For comparison the values obtained from \cite{rang95} are shown as filled squares.
As in Figure \ref{model_prof} the vertical dotted line marks the position of the PSPC support ring.}
\label{density_prof}
\end{figure}
%#############################################

We have derived the hot gas density radial profile of NGC 1399 from the X-ray surface brightness profile, making use of the procedure described by \cite{kriss83}, that allows to deproject the observed emission in concentric shells, assuming spherical symmetry and homogeneity of the hot gas. Since the global gas distribution has a complex and asymmetric structure, we cannot apply this method to the total emission; instead we applied it to the `central' and `galactic' components separately, whose ellipticities, derived from the bidimensional models (see previous section), are small.
The spherical symmetry approximation is not valid for the cluster halo so that, in deriving the density profile, we assumed rotational symmetry around the minor axis. We found that assuming symmetry around the major axis the difference is of the order of a few percents (except at very large radii where exposure correction uncertainties are  dominant) and resulted in minor changes to our conclusions. The central, galactic and cluster components were assumed to be isothermal with temperatures of respectively kT=0.86, 1.1 and 1.1 keV, in agreement with the PSPC temperature profiles derived by RFFJ. \cite{Buo99} showed that ASCA data suggest the presence of multiphase gas within the central 5' of NGC 1399, better fitted by cooling flows models. However, as he notes, the narrower ROSAT energy range and the lower spectral resolution is unable to clearly discriminate between single and multi-temperature models (even though it reveals the presence of radial temperature gradients) so that our assumption of isothermal components is adequate for the kind of analysis performed here. 

The electron density profiles of the isothermal components are shown in Figure \ref{density_prof}. The density profiles are centered on the centers of their respective components and thus the total density, shown as a thick line, does not take into account the centers offset. Including the displacements of each component with respect to the galaxy center results in a slightly smoother profile in the outer regions, due to the azimuthal averaging,  shown as a thin continuous line. ROSAT and ASCA spectral fits (RFFJ; \citealp{Buo99,Matsu00}) revealed that additional absorption, over the galactic value, may be required in the galaxy center. We thus estimated an upper limit for the density of the central component assuming an absorption of $2.0\times 10^{21}$ cm$^{-2}$ (see RFFJ), shown as downward triangles in Figure \ref{density_prof}.

We must notice that adding the single density components, derived from the surface brightness decomposition, to obtain the total density profile is not strictly the correct approach since the surface brightness does not depend linearly on the gas density. However regions of the image in which more than one component are significant are small and the good agreement of our results with the estimate of RFFJ (who deprojected the total surface brightness) indicates that we are not introducing significant errors in our total density estimate. The same considerations hold for the mass profiles derived below.

%#############################################
\begin{figure}[th]
\centerline{\psfig{figure=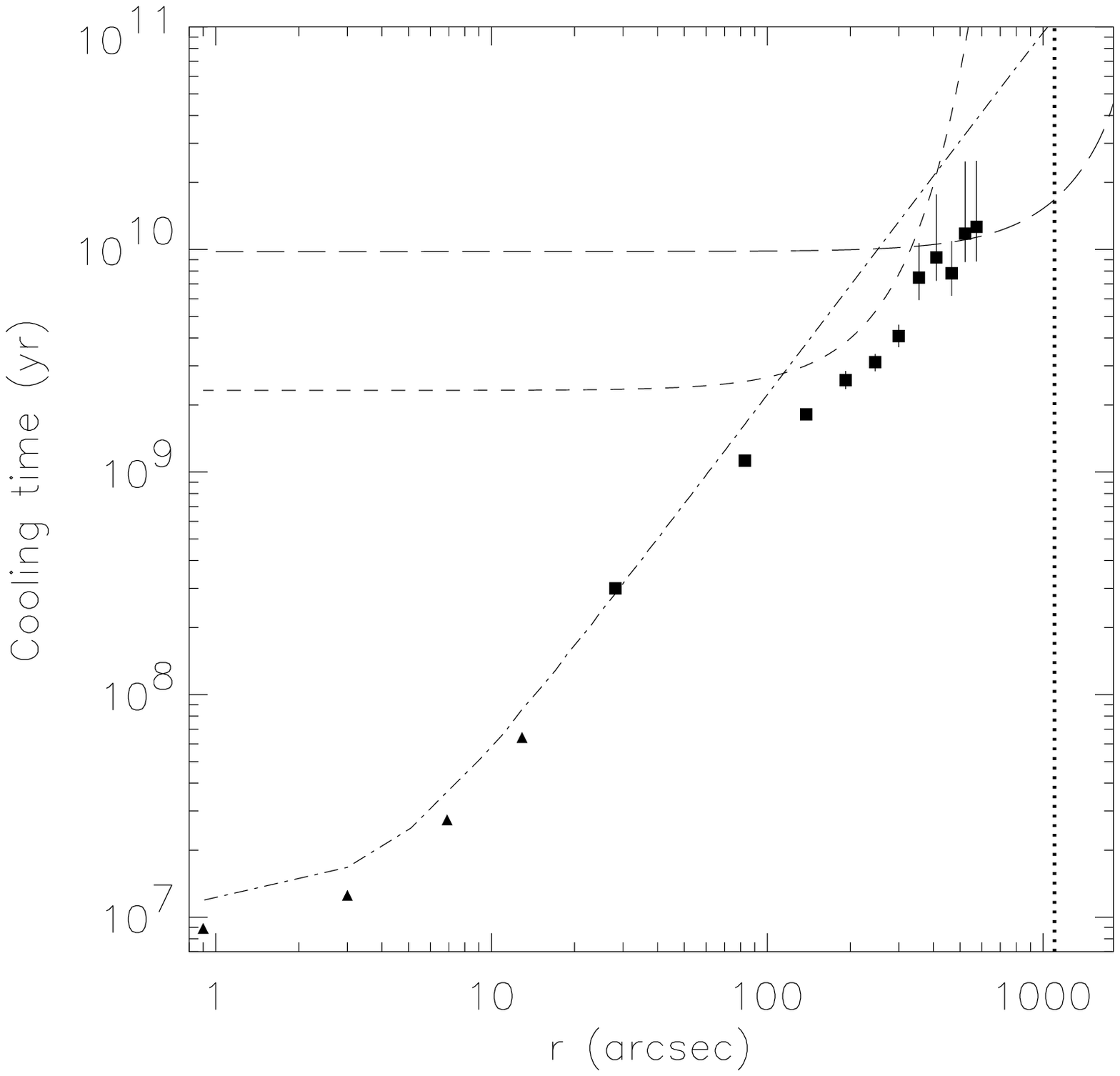,angle=0,width=0.35\textheight}}
\caption{Cooling time profiles derived from deprojection. The symbols have the same meaning as in Figure \ref{density_prof}. Note that the assumption of a high central absorption translates into lower limits on the central cooling time (filled triangles).}
\label{cooling_prof}
\end{figure}
%#############################################

Using the density profiles and the assumed temperatures we estimated the cooling time $\tau_c\propto nkT/n_e n_H \Lambda$ as a function of radius, where  $n_e$ is the electron density shown in Figure \ref{density_prof}, $n_H=n_e/1.21$ \citep{Sar88} and $n=n_e+n_H$ are the hydrogen and total density and $\Lambda$ is the cooling function given by \cite{Sar87}. Figure \ref{cooling_prof} shows that the central and galactic components have cooling times much shorter than the age of the universe, assumed to be $\sim 10^{10}$ years, out to 250'' and 350'' (23 and 32 kpc) respectively.

%#############################################
\begin{figure*}[th]
\subfigure[]{\psfig{figure=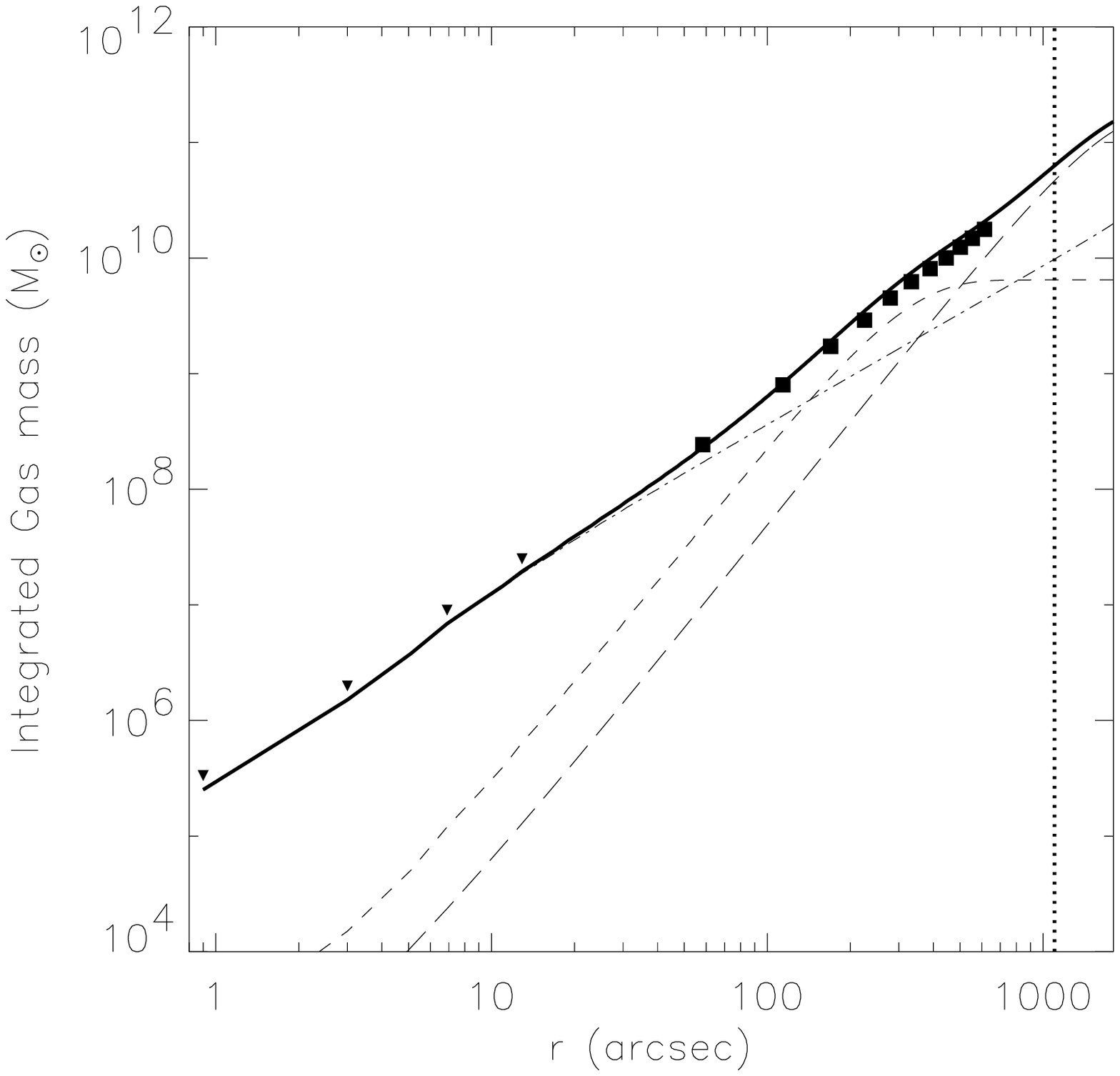,angle=0,width=0.45\textwidth}}
\subfigure[]{\psfig{figure=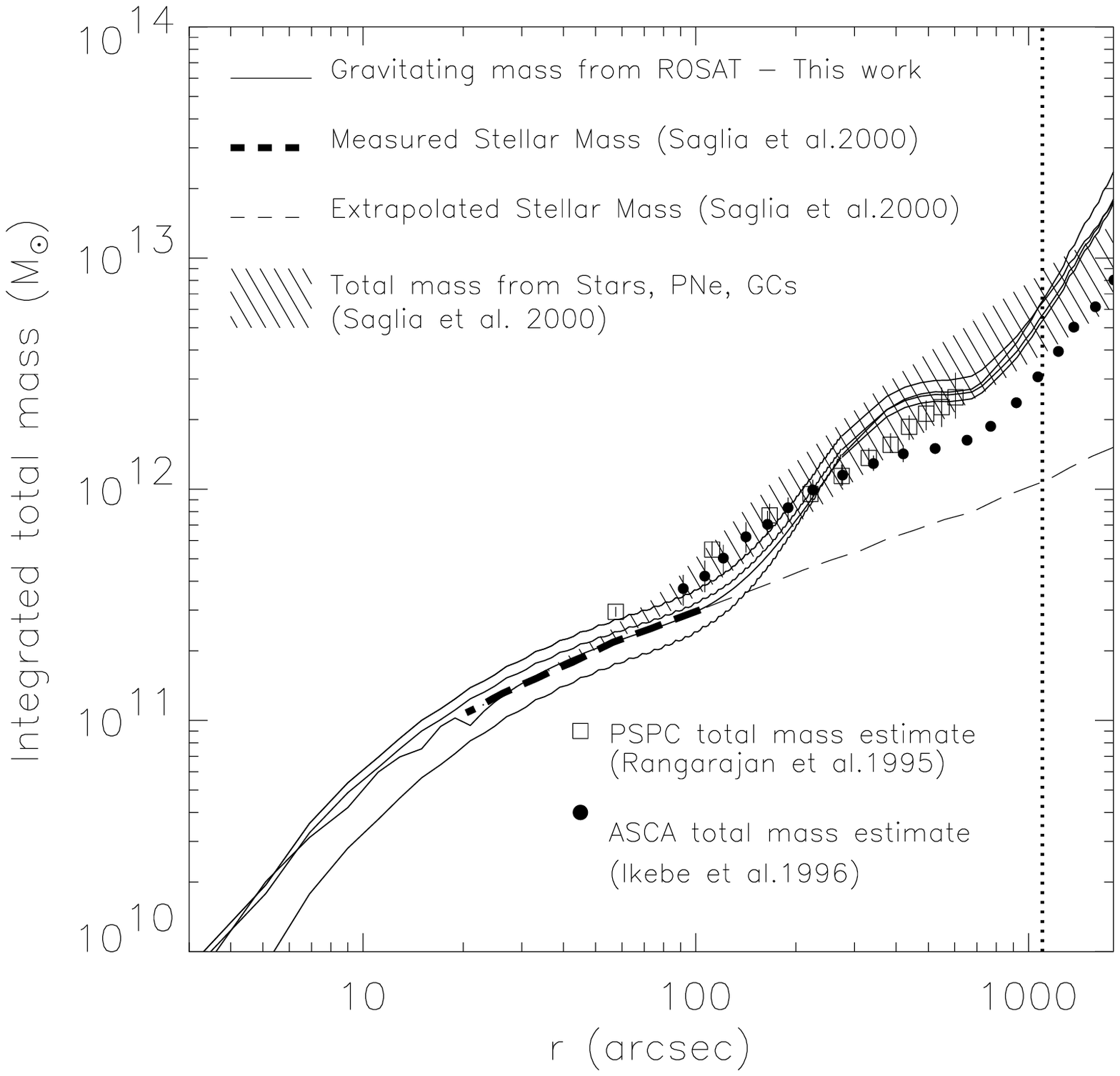,angle=0,width=0.45\textwidth}}
\caption{Integrated mass profiles derived from deprojection. {\bf (a)} Integrated gass mass, with symbols having the same meaning as in Figure \ref{density_prof}. {\bf (b)} Integrated gravitating mass for our different temperature profiles (see discussion in text) are shown as continuous lines. Open squares and filled circles represent the total gravitating mass estimated respectively by \cite{rang95} and \cite{ikebe96}. The gravitating mass range derived by \cite{saglia00} from optical observations of stars, GCs and PNe, is shown as a shaded region. The contribution of the stellar matter measured (extrapolated) by \cite{saglia00} is represented by the thick (thin) dashed line.}
\label{mass_prof}
\end{figure*}
%#############################################

The cumulative gas mass profile is shown in Figure \ref{mass_prof}a as a function of radius. The flat profile of the galactic component over 600'' (55 kpc) reflects the steep slope of the Beta profile. The filled triangles show that the adoption of a high absorbing column in the galaxy center does not affect significantly the gas mass estimates, giving a difference of $\sim $33\%. 
We estimated the total gravitating mass using the expression \citep{Fabr80}:
\begin{equation}
\label{mass_eq}
M(<r)=-\frac{kT(r)}{G\mu m_p}\left( \frac{d\ln\rho}{d\ln r}+\frac{d\ln T}{d\ln r}\right) r
\end{equation}
We tried four different temperature profiles based on published data: we used both the linear and power-law approximations derived by \cite{jones97} based on PSPC spectral fits, a constant temperature of 1.1 keV that approximates the isothermal profile found by RFFJ and a power-law profile $T(r)=T_0 r^\alpha$,
with $T_0=0.6$ keV and $\alpha=0.13$, representing the temperature drop in the inner 150" (14 kpc). The gravitating mass, extracted in concentric annuli centered on the X-ray centroid (i.e. taking into account the offset between the different components), is shown in Figure \ref{mass_prof}b. The similarity of the continuous lines indicates that the gravitating mass is weekly dependent on the assumed temperature profile and is mostly determined by the density profile. The largest deviation is seen in the model taking into account the central temperature drop, which is systematically lower in the inner 100". 
Our profiles are in good agreement with RFFJ and fall within the range predicted by optical observations (shaded region, \citealp{saglia00}). 
We clearly distinguish the transition between the galaxy and cluster halo near 700" (64 kpc), in agreement with the ASCA data (filled circles, \citealp{ikebe96}). Both components exceed the luminous mass, based on an extrapolation of stellar measurements (thin dashed line), indicating the presence of a consistent amount of dark matter.

Thanks to the ROSAT HRI resolution in the inner 100" (9 kpc) we are able to resolve the dynamical behavior of the central component. The gravitating mass follows closely the stellar mass profile, confirming that the central galaxy dynamics are dominated by the luminous matter \citep{saglia00}.    

\subsubsection{NGC 1404}
\label{dens_par_1404} 
\begin{figure*}[t!]
\subfigure[]{\psfig{figure=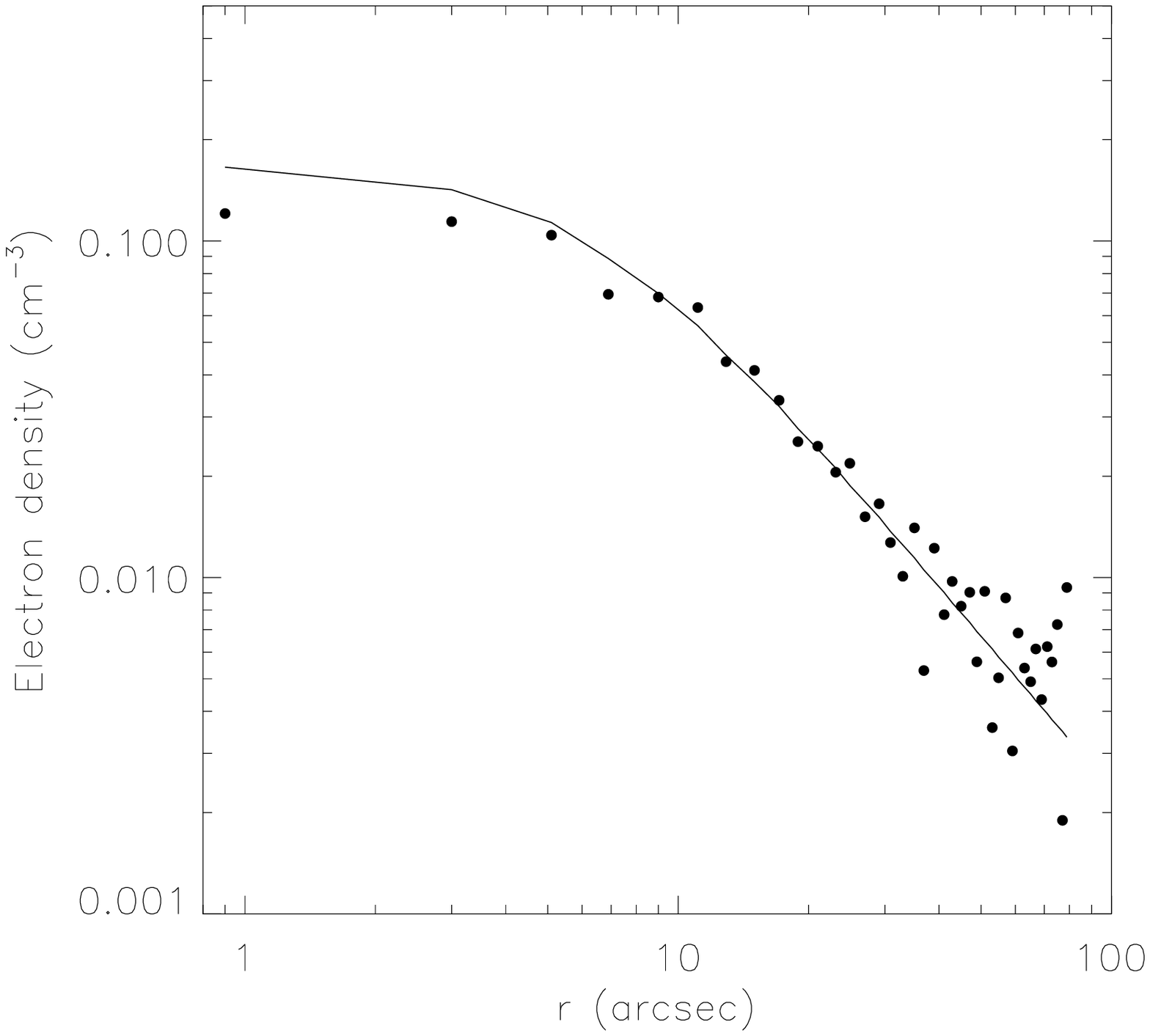,angle=0,width=0.45\textwidth}}
\subfigure[]{\psfig{figure=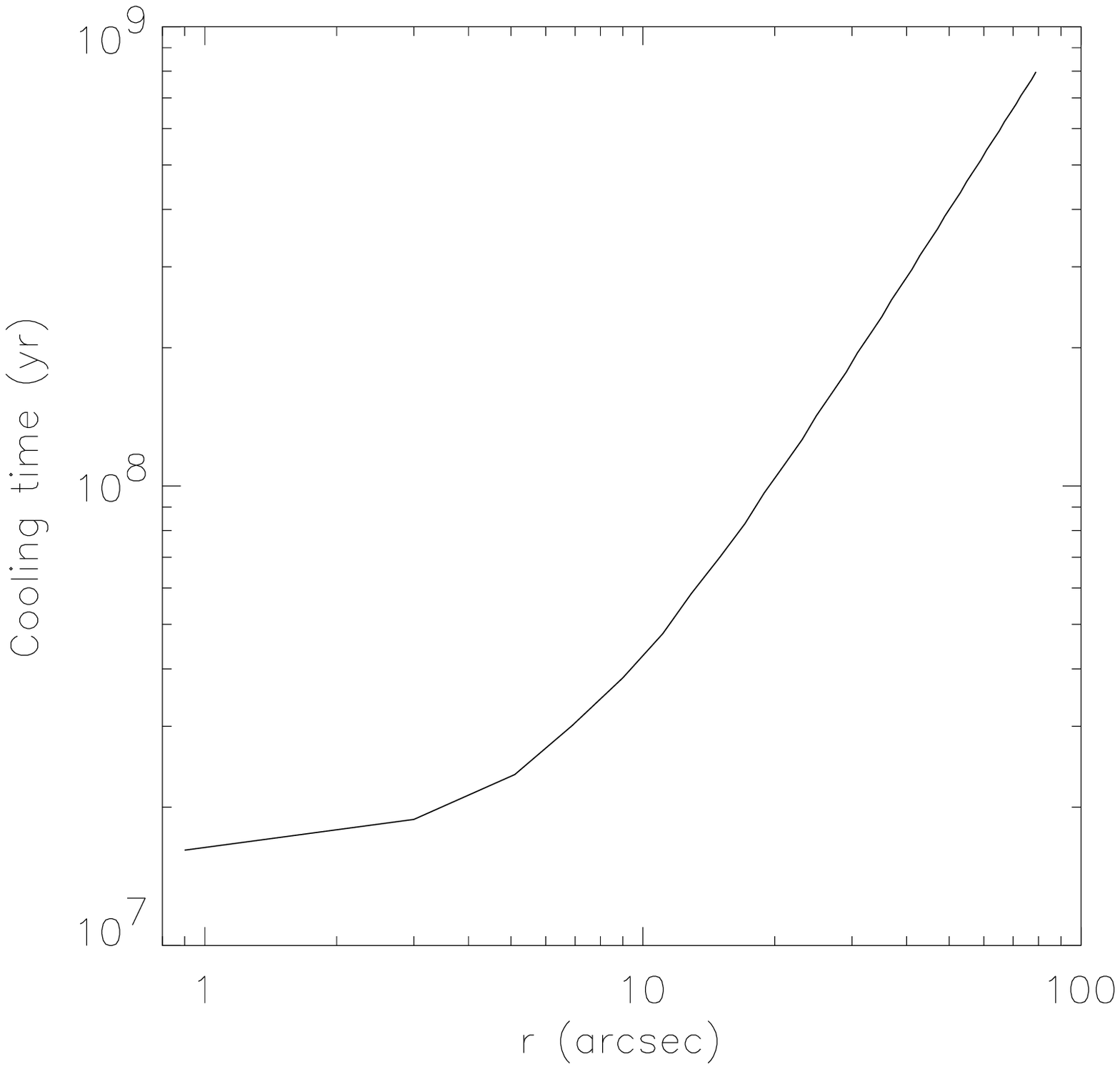,angle=0,width=0.45\textwidth}}
\caption{{\bf (a)} Deprojected electron density profile of NGC 1404. The filled circles represent the profile obtained from direct surface brightness deprojection while the continuous line shows the smoother profile obtained by deprojecting the best-fit Beta model. {\bf (b)} Cooling time profile derived from the best fit Beta model.}
\label{density_prof_1404}
\end{figure*}

Following the method outlined in the previous section we derived the electron density, cooling time and mass profiles for NGC 1404. Since the deprojection 
algorithm tends to enhance the fluctuations present in the brightness profile,
we used the best fit model to derive the deprojected electron density. This is possible because the NGC 1404 radial brightness profile was well fitted by a single symmetric Beta model out to 90" ($\S$ \ref{NGC 1404}). 
Figure \ref{density_prof_1404}a shows the good agreement between the deprojected density derived from the Beta model (continuous line) and the one from the surface brightness profile (filled circles). The discrepancy in the inner 5" is due to the fact that the Beta model is deconvolved for the instrumental PRF. 
We stopped the deprojection at 80" to avoid uncertainties related to the background subtraction and to the asymmetries due to ram pressure stripping ($\S$ \ref{NGC 1404}), both of which tend to increase the fluctuations at large radii.

%#############################################
%#############################################

The cooling time profile derived from the Beta model is shown in Figure \ref{density_prof_1404}b and suggests that significative cooling must be present in the NGC 1404 halo because the cooling time is smaller than $10^9$ yr within the inner 80" ($\sim$7 kpc). We adopted a cooling function $\Lambda=3.8\times10^{-23}$ ergs cm$^3$ s$^{-1}$ for kT=0.6 keV \citep{Sar87}.

 %#############################################
\begin{figure}[t!]
\centerline{\psfig{figure=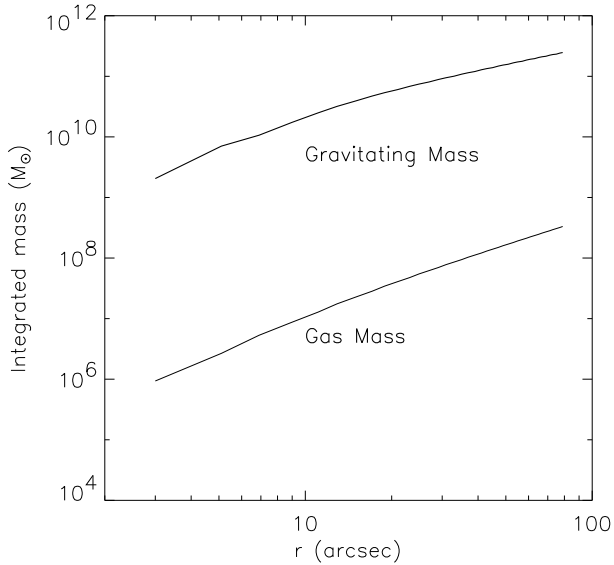,angle=0,width=0.4\textheight}}
\caption{Integrated mass profiles of NGC 1404 derived from deprojection. The total gas mass is represented by the lower line while the total mass derived using equation \ref{mass_eq} is shown by the upper line.}
\label{mass_prof_1404}
\end{figure}
%#############################################

The integrated gas mass and the total gravitating mass derived from equation \ref{mass_eq} are shown in Figure \ref{mass_prof_1404}. We assumed NGC 1404 to have an isothermal profile with kT=0.6 keV \citep{jones97}.

\subsection{Global Fluxes and Luminosities}
\label{fluxes}

To determine the total HRI X-ray flux of NGC 1399 we measured the total net counts within a 500'' radius from the X-ray centroid. The background counts were extracted from the rescaled particle map (see $\S$ \ref{NGC 1399}) in the 500''--850'' annulus .
We then corrected the background subtracted counts for the contamination of pointlike sources within the 500'' circle. This was done by subtracting, for each source,  the net counts measured by the wavelets algorithm (see $\S$ \ref{sources}). The final count rate, for the hot halo of NGC 1399, is 0.451$\pm$0.004 counts sec$^{-1}$.

To convert this count rate into a flux we used the best fit Raymond-Smith (RS) model found by RFFJ from the ROSAT PSPC data (a main thermal component with  $kT\simeq$1 keV plus a softer emission with $kT=80.8$ eV contributing for 16\% of the emission in the 0.2--0.3 keV range). Using the HEASARC-XSPEC software, we simulated a RS spectrum fixing the absorbing column to the galactic value ($1.3 \times 10^{20}$ cm$^{-2}$) and the metal abundance to the solar value. With this spectrum, we calculated the conversion factor between counts and fluxes using the PIMMS software, that takes into account the HRI spectral response function. We obtain 1 count sec$^{-1}$=3.158$\times 10^{-11}$ erg s$^{-1}$ cm$^{-2}$ so that $f_X=(1.42\pm 0.01)\times 10^{-11}$ erg s$^{-1}$ cm$^{-2}$ in the 0.1-2.4 keV energy range. Assuming a distance of 19 Mpc, $L_X(0.1-2.4 keV)=(5.50\pm 0.04)\times10^{41}$ erg s$^{-1}$. Taking into account the uncertainties in the exposure correction and background extraction, this result agrees within 10\% with the estimate of RFFJ, converted to our adopted distance.

In the case of NGC 1404 we used a single temperature Raymond-Smith spectrum with $kT=0.6$ keV and metal abundance 0.2 solar, following \cite{jones97}. We measured a count rate of $0.117\pm0.001$ counts sec$^{-1}$ within a 150'' radius, extracting the background from the 200''-300'' annulus. The corresponding (0.1-2.4 keV) flux and luminosity are $f_X=(4.06\pm 0.03)\times 10^{-12}$ erg s$^{-1}$ cm$^{-2}$ and $L_X=(1.57\pm 0.01)\times10^{41}$ erg s$^{-1}$.

We also searched for evidence of nuclear variability of the two galaxies in our data, finding none.

\subsection{X-ray/Radio Comparison}
\label{Xradio}
\begin{figure*}[t]
\centerline{\psfig{figure=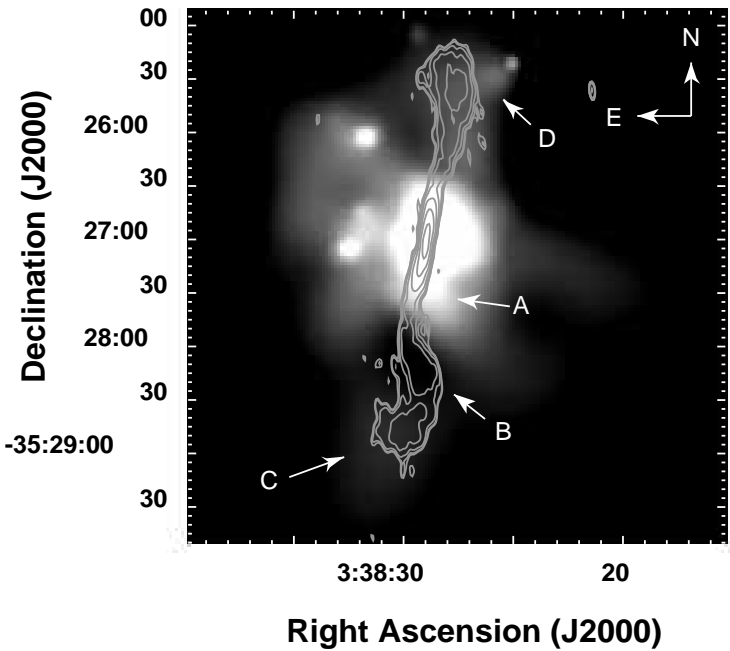,angle=0,width=1.\textwidth}}
\caption{6 cm radio contours (green; from Figure 1 of \citealp{kbe88}) superimposed on the adaptively smoothed 1''/pixel X-ray image.
The logarithmic grayscale represents intensities from 8.6$\times 10^{-3}$ (black) to 6.7$\times 10^{-1}$ (white) cnts arcmin$^{-2}$ s$^{-1}$.}
\label{X-radio}
\end{figure*}

NGC 1399 is known to be a radio galaxy with a faint radio core ($S_{core}\sim 10$ mJy at 5 GHz, \citealp{ekers89}). \cite{kbe88} studied in detail the 6 and 20 cm radio emission. The 6 cm radio contours (their Figure 1) are superimposed on the `csmoothed' X-ray image in Figure \ref{X-radio}. The radio core lies in the center of the galaxy, but the nuclear source is not distinguishable from the radio jets at this resolution.

We searched our X-ray data for a central point source that could be the counterpart of the radio core. Trying to model the nuclear emission with a Beta model plus a delta function, convolved with the HRI PRF, resulted in minor variations in the Beta model parameters and no improvement in the fit statistics, thus indicating no need for additional nuclear emission. To estimate an upper limit we varied the central point source brightness and repeated the fit until we reached a $3\sigma$ confidence level. We obtain an upper count rate limit of $2.25\times 10^{-3}$ counts s$^{-1}$. Assuming a power law spectrum with $\alpha_{ph}=1.7$ and galactic absorption, we get $f_X^{3\sigma}=1.0\times 10^{-13}$ erg s$^{-1}$ cm$^{-2}$ in the 0.1-2.4 keV band, corresponding to $L_X^{3\sigma}=3.9\times 10^{39}$ erg s$^{-1}$. This value is approximately half of the one derived by \cite{Sulk01} making use of one of our HRI observations (RH600831a01) and 30\% of the limit estimated by RFFJ by means of spectral analysis. Recently \cite{Loew01} were able to further reduce the upper limit on the nuclear source by one order of magnitude making use of {\it Chandra} data.

%#############################################
%#############################################
    
By comparing the 6 cm radio contours with the X-ray halo (Figure \ref{X-radio}) we identify several interesting features:
i) in the X-rays the nuclear region is elongated in the N-S direction, following the radio jets. This elongation is more pronounced in the Southern direction, where there is an excess of X-ray emission following the Western side of the jet (region A); ii) At the point where the Southern jet ends in the lobe there is a steep gradient in the X-ray emission. The lobe is bent towards the West and aligned with a region of low X-ray emission (region B); iii) The Southern lobe ends in an irregular clump coincident with a region of higher X-ray emission (region C); iv) No features as the ones seen in the Southern lobe are evident in the Northern one. Here the radio jet ends smoothly in a regular lobe with no sign of sharp transition and there is no evidence of cavities (minima) in the X-ray emission. Instead the jet and the lobe are aligned with an X-ray structure that seems to be coincident with the right side of the jet/lobe. At the North-West end of the Northern lobe there are two X-ray clumps (region D): they are detected as a single source by the wavelet algorithm (No.32 in Table \ref{source_tab}). To a more accurate visual inspection we found that the smaller and compact source is coincident with an optical (probably background) counterpart, while there is no counterpart for the diffuse clump ($\S$ \ref{sources}) suggesting that it is a local feature of the hot gaseous halo.

From the residual image of the two-dimensional fit (Figure \ref{residuals}) we can calculate the significance of these features. We find that the depletion corresponding to the Southern lobe (region B) and the Northern diffuse clump (region D) are significant respectively at the $2.5\sigma$ and $\sim 4\sigma$ level. Region C corresponds to a $3\sigma$ brightness enhancement ($\S$ \ref{model}).

%#############################################
\begin{figure}[t]
\centerline{\psfig{figure=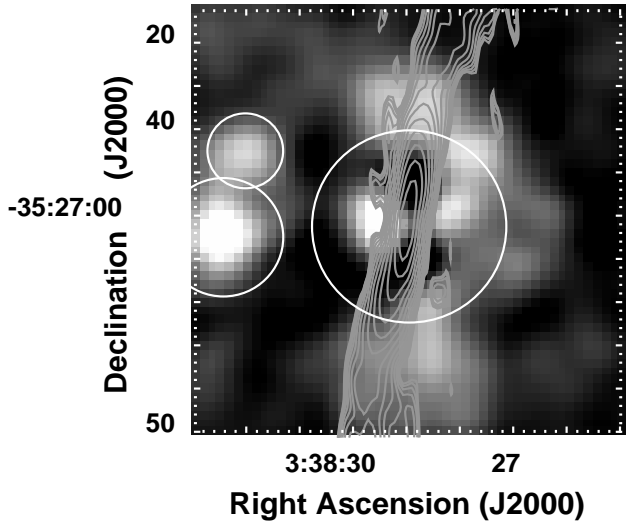,angle=0,width=0.6\textwidth}}
\caption{6 cm radio contours superimposed on the 2''/pixel residuals X-ray image of the nuclear region. The image has been smoothed with a gaussian of $\sigma= 4"$. The logarithmic grayscale represents residuals ranging from $-2.2 \sigma$ (black) to $+4 \sigma$ (white). 
The circles on the left show the positions of X-ray sources detected by the wavelets algorithm (No.2 and 11 in Table \ref{source_tab}). The radius of the central source (corresponding to the central X-ray peak) is $\sim 22"$.}
\label{X-radio_nuclear}
\end{figure}
%#############################################

To examine in greater detail the nuclear region we subtracted the halo model developed in $\S$ \ref{model} from the X-ray image to produce a residual map of the nuclear region with a 2"/pixel resolution (Figure \ref{X-radio_nuclear}).
As discussed at the beginning of this section, there is no evidence of a nuclear point source aligned with the radio centroid. Instead residual emission is visible on both sides of the radio jets a few arcseconds ($\leq 1$ kpc) from the nucleus. There is no evidence that any of these excesses may be due to the nuclear source because they are displaced with respect to the radio emission. Moreover the radio emission seems to avoid the X-ray excess on the eastern side of the nucleus, suggesting that these features may be correlated. 

In coincidence with the direction of the radio jets there are two regions of negative residuals ($\sim -2\sigma$), extending up to $\sim 20"$, where
additional excess emission, on both the Northern and Southern sides of the X-ray core and aligned with the radio jets, is clearly visible. The Southern excess corresponds to region A of Figure \ref{X-radio}.
Residual emission is also present on the Western side of the nuclear region. 
Visual inspection of the {\it Chandra} image (see $\S$ \ref{Chandra}) shows
that this excess is probably due to the presence of point sources unresolved in the HRI image.

%#############################################
\begin{figure}[t]
\centerline{\psfig{figure=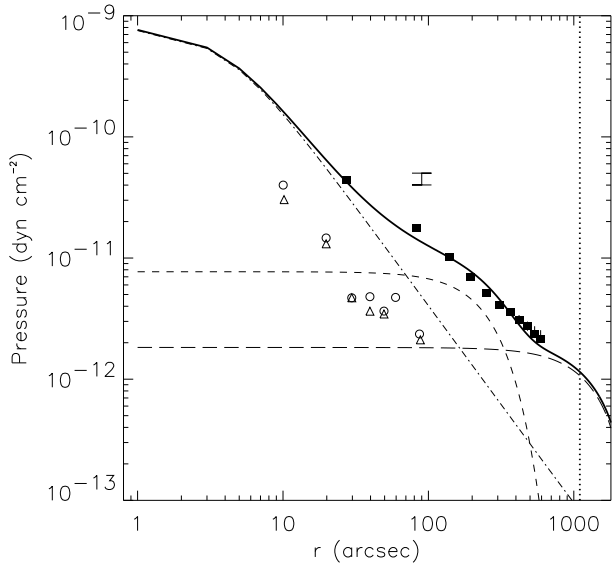,angle=0,width=0.4\textheight}}
\caption{Gas pressure profiles derived from deprojection. The symbols have the same meaning of Figure \ref{density_prof}. Open circles and triangles represent respectively the radio pressure of the Northern and Southern lobes measured by \cite{kbe88}. The vertical error bar in the Figure indicates the pressure range of the D region (see discussion in text).}
\label{press_prof}
\end{figure}
%#############################################

From the density profiles obtained in $\S$ \ref{dens_par} we were able to obtain the pressure profiles of the three halo components shown in Figure \ref{press_prof}. The pressure profiles are compared to the minimum radio pressure measured by \cite{kbe88} in the Northern (circles) and Southern (triangles) radio jets/lobes. The Figure confirms their result that the radio source can be confined by the ISM whose pressure is higher than the radio one. We must notice that they found a difference between radio and thermal pressure of more than one order of magnitude, while we find $P_{Thermal}/P_{Radio}= 3 \div 4$. This is mostly due to the poor ISM temperature constraint obtained from the {\it Einstein} IPC data and to our better spatial resolution. 

The vertical error bar in the Figure indicates the pressure estimate for the diffuse clump observed in the D region. To calculate this value we assumed that the brightness enhancement is due to a homogeneous sphere of $\sim 9"$ (800 pc) radius at the same temperature of the galactic halo (1.1 keV). Depending on whether the background is extracted locally (an annulus 9" wide) or from the Beta model, the `bubble' pressure can vary up to 25\%, so that the allowed range is shown as an error bar.
This estimate is dependent on the assumed geometry of the emitting region and on projection effects, so that it must be considered as an upper limit on the actual pressure.

\subsection{Discrete Sources}
\label{sources}
%#######################################################################
\begin{figure*}[!t]
\centerline{\psfig{figure=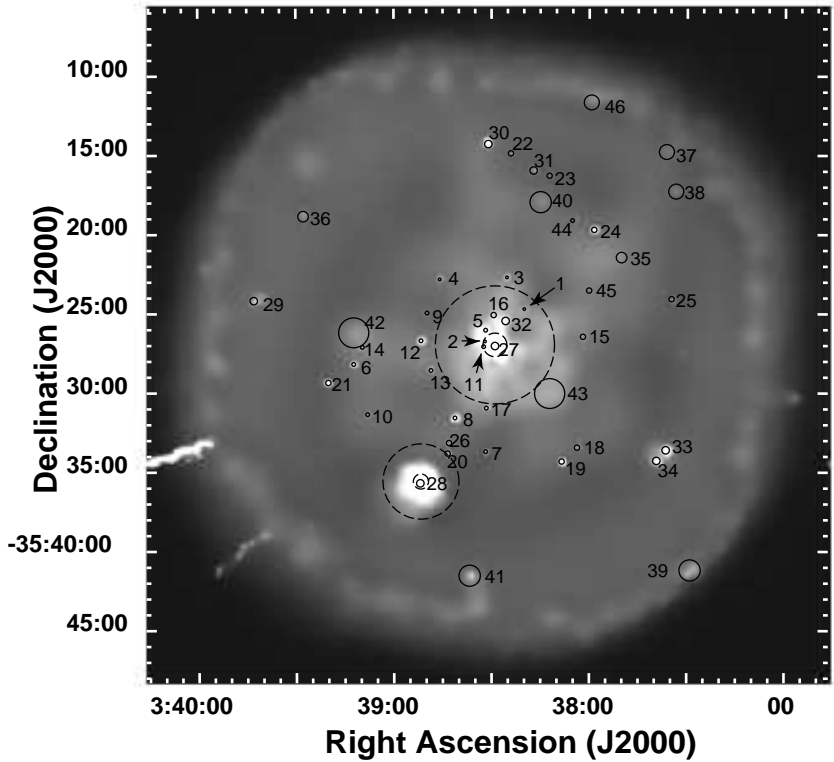,angle=0,width=0.8\textwidth}}
\caption{The sources detected with the wavelets algorithm of \cite{Dam97a,Dam97b} superimposed on the 5''/pixel adaptively smoothed HRI image. Solid circles show the region of maximum S/N ratio for each source. Dashed circles represent $r_{eff}$ and $5r_{eff}$ for NGC 1399 and NGC 1404 \citep{gou94}.}
\label{det}
\end{figure*}
%#######################################################################
We used the algorithm developed at the Palermo Observatory by F. Damiani and collaborators \citep{Dam97a} to detect the discrete sources present in the HRI field. The algorithm measures the relevant parameters in 
a wavelet transformed space, using different wavelet scales to optimize the detection of both compact and extended sources.  
The wavelet algorithm found 43 sources in our field, using a signal to noise threshold of 4.65. The threshold was chosen to assure a contamination of one/two spurious detections per field (F. Damiani private communication, see also \citealp{Dam97b}). Three more sources were found with the algorithm, but being clearly associated to the ribbon-like features due to the presence of hot spots in the detector (Figure \ref{NGC1399comp}), were excluded from the subsequent analysis. 
After visual inspection of the adaptively smoothed image (Figure \ref{csmooth}) three more detections (No.44, 45, 46) with a lower S/N ratio but showing clear features of pointlike sources, were added to the list.

The source positions in the field are shown in Figure \ref{det}. Sources No.42 and 43 are clearly extended and seem to be associated to structures in the NGC 1399 halo revealed by the adaptive smoothing algorithm ($\S$ \ref{brightness}).

Table \ref{source_tab} lists the parameters measured with the wavelets algorithm for each source: position, 3$\sigma$ radius, number of counts with statistical error, maximum signal to noise ratio (for sources detected at different spatial scales), count rate and flux in the 0.5-2.0 keV band for both HRI and PSPC data. The 0.5-2.0 keV energy band
was chosen to allow an easily comparison with the \cite{Has93} background counts (see below). 
The the last three columns of Table \ref{source_tab} report the source short-term variability obtained with a Kolmogorov-Smirnov test of the HRI light curves, long-term variability from comparison with PSPC fluxes (as explained below),
and possible optical counterparts (within the 3$\sigma$ radius measured by the wavelets algorithm) found by both visual inspection of the DSS photographic plates and cross-checking with known sources present in the Nasa Extragalactic Database (NED). The names of the counterparts correspond to those reported by \cite{Hilk99a} or \cite{Smith96} in their optical and NUV studies of the Fornax field.
Sources No.27 and 28, representing respectively NGC 1399 and NGC 1404, are not included in the table.

%#######################################################################
\begin{landscape}
\topmargin=1.in
\begin{table}[p]
\caption{Discrete Sources Properties\label{source_tab}}
\tiny
\begin{tabular}{cccrrrccccccc}
\\
\tableline
\tableline
Source & R.A.         & Dec.         & 3$\sigma$ Radius & Counts & Max S/N & Count Rate                   & $\mathit{f}_{HRI (0.5-2.0~keV)}^{pow.law}$ & $\mathit{f}_{HRI (0.5-2.0~keV)}^{max~abs.}$ & $\mathit{f}_{PSPC (0.5-2.0~keV)}^{pow.law}$ & Time Var.\tablenotemark{a} & Opt.id. & NED obj.\tablenotemark{b}\\
 No.  & \multicolumn{2}{c}{(J2000)} & (arcsec)         &        &         & ($10^{-4}$ Cnts sec$^{-1}$) & ($10^{-15}$ erg s$^{-1}$ cm$^{-2}$)   & ($10^{-15}$ erg s$^{-1}$ cm$^{-2}$) &  ($10^{-15}$ erg s$^{-1}$ cm$^{-2}$) & short-term / long-term &         & \\
\tableline
1 & 3:38:20.0 	  & -35:24:42    & 8.7      & 35$\pm$12   & 4.7  & 2.1$\pm$0.7  & 4.3$\pm 0.2$ 	 &   5.6 & 28$\pm 7$ 	 & n/y & - & - \\
	
2 & 3:38:32.0 	  & -35:26:45    & 8.7      & 54$\pm$17   & 6.1  & 3.2$\pm$1.0  & 6.7$\pm 0.3$ 	 &   8.6 & - 	 & n/$\cdots$ & - & - \\
	
3 & 3:38:25.3 	  & -35:22:42    & 8.7      & 49$\pm$14   & 6.7  & 2.9$\pm$0.8  & 6.1$\pm 0.3$ 	 &   7.8 & - 	 & n/$\cdots$ & - & - \\
	
4 & 3:38:45.9 	  & -35:22:50    & 8.7      & 42$\pm$12   & 6.1  & 2.6$\pm$0.7  & 5.3$\pm 0.3$ 	 &   6.9 & 9$\pm 3$ & n/n & y & CGF 0202\\
	
5 & 3:38:31.8 	  & -35:26:02    & 11.9     & 165$\pm$23  & 13.0 & 10$\pm$2  & 20$\pm 1$	 &  26.0 & -	 & n/$\cdots$ & - & CGF 0233\\
	
6 & 3:39:12.3 	  & -35:28:12    & 13.2     & 172$\pm$21  & 14.9 & 10$\pm$1 & 21$\pm 1$	 &  27.5 & 25$\pm 4$	 & n/n & y & - \\
	
7 & 3:38:31.8 	  & -35:33:42    & 12.2     & 56$\pm$16   & 6.0  & 3.3$\pm$0.9  & 6.9$\pm 0.3$	 &   8.9 & 8$\pm 3$	 & n/n & - & - \\
	
8 & 3:38:41.2 	  & -35:31:35    & 10.8     & 1000$\pm$35 & 60.6 & 59$\pm$2 & 123$\pm 6$     & 159.2 & 170$\pm 8$ & 99\%/y & y & - \\
	
9 & 3:38:49.8 	  & -35:24:57    & 12.2     & 47$\pm$14   & 5.0  & 2.8$\pm$0.9  & 5.8$\pm 0.3$ 	 &   7.5 & - 	 & n/$\cdots$ & - & - \\
	
10 & 3:39:08.1	  & -35:31:22    & 12.2     & 56$\pm$16   & 6.2  & 3.3$\pm$0.9  & 6.9$\pm 0.4$ 	 &   8.9 & 22$\pm 6$ 	 & n/n & - & - \\
	
11 & 3:38:32.4	  & -35:27:05    & 13.7     & 171$\pm$30  & 11.5 & 10$\pm$2 & 21$\pm 1$	 &  27.2 & -	 & n/$\cdots$ & - & - \\
	
12 & 3:38:51.7	  & -35:26:42    & 11.6     & 295$\pm$22  & 24.7 & 17$\pm$1 & 36$\pm 2$	 &  47.0 & 58$\pm 6$	 & 95\%/y & y & CGF 0102\\
	
13 & 3:38:48.6	  & -35:28:35    & 12.2     & 77$\pm$20   & 7.7  & 4.5$\pm$1.1  & 9.3$\pm 0.5$ 	 &  12.0 & - 	 & n/$\cdots$ & y & - \\
	
14 & 3:39:09.7	  & -35:27:07    & 12.2     & 46$\pm$14   & 5.0  & 2.8$\pm$0.8  & 5.7$\pm 0.3$ 	 &   7.4 & - 	 & n/$\cdots$ & y & - \\

15 & 3:38:01.9    & -35:26:27    & 17.3     & 62$\pm$19   & 4.7  & 3.6$\pm$1.1  & 7.5$\pm 0.4$ 	 &   9.7 & - 	 & n/$\cdots$ & - & - \\
	
16 & 3:38:29.4    & -35:25:05    & 17.3     & 75$\pm$23   & 4.8  & 4.4$\pm$1.3  & 9.1$\pm 0.5$ 	 &  11.7 & - 	 & n/$\cdots$ & - & - \\
	
17 & 3:38:31.6    & -35:30:58    & 12.2     & 70$\pm$18   & 6.8  & 4.1$\pm$1.1  & 8.6$\pm 0.5$ 	 &  11.1 & - 	 & n/$\cdots$ & - & - \\
	
18 & 3:38:03.7    & -35:33:27    & 17.3     & 65$\pm$19   & 5.4  & 4.1$\pm$1.2  & 8.5$\pm 0.4$ 	 &  10.9 & - 	 & n/$\cdots$ & y & CGF 0354\\
	
19 & 3:38:08.5    & -35:34:20    & 15.0     & 908$\pm$34  & 52.3 & 57$\pm$2	 & 119$\pm 6$    & 152.7 & 127$\pm 7$ & 99\%/n & y & - \\

20 & 3:38:43.5    & -35:33:50    & 15.2     & 105$\pm$18  & 8.7  & 6.3$\pm$1.1  & 13.1$\pm 0.7$	 &  16.8 & -	 & n/$\cdots$ & y & - \\

21 & 3:39:20.1    & -35:29:22    & 16.5     & 527$\pm$30  & 33.0 & 32$\pm$2	 & 68$\pm 4$  &  86.9 & 83$\pm 7$  & 99\%/n & y & - \\
 
22 & 3:38:24.1    & -35:14:52    & 17.3     & 60$\pm$18   & 4.9  & 3.8$\pm$1.2  & 8.0$\pm 0.4$ 	 &  10.3 & 12$\pm 4$ 	 & n/n & y & - \\

23 & 3:38:12.2    & -35:16:17    & 17.3     & 59$\pm$18   & 4.8  & 3.7$\pm$1.1  & 7.7$\pm 0.4$ 	 &   9.9 & - 	 & n/$\cdots$ & y & - \\
%\tableline
%\end{tabular}
%\tablenotetext{a}{\small Short-term = probability of the Kolmogorov-Smirnov test: `n' means no variability detected at the 95\% level.\\
%		  Long-term = difference between PSPC and HRI counts: `y' means variable source ($>3\sigma$), `n' means no variability detected, `$\cdots$' means no available data.}
%\tablenotetext{b}{\small CGF=\cite{Hilk99a}; FCCB=\cite{Smith96}}
%\end{table}
%\end{landscape}
%#######################################################################
%#######################################################################
%\begin{landscape}
%\topmargin=1.in
%\thispagestyle{empty}
%\begin{table}[p]
%\addtocounter{table}{-1}
%\caption{ - Continued}
%\tiny
%\begin{tabular}{cccrrrcccccccc}
%\\
%\tableline
%\tableline
%Source & R.A.         & Dec.         & 3$\sigma$ Radius & Counts & Max S/N & Count Rate                   & $\mathit{f}_{HRI (0.5-2.0~keV)}^{pow.law}$ & $\mathit{f}_{HRI (0.5-2.0~keV)}^{max~abs.}$ & $\mathit{f}_{PSPC (0.5-2.0~keV)}^{pow.law}$ & Time Var.\tablenotemark{a} & Opt.id. & NED obj.\tablenotemark{b}\\
%No.   & \multicolumn{2}{c}{(J2000)} & (arcsec)         &        &         & ($10^{-4}$ Cnts sec$^{-1}$) & ($10^{-15}$ erg s$^{-1}$ cm$^{-2}$)   & ($10^{-15}$ erg s$^{-1}$ cm$^{-2}$) &  ($10^{-15}$ erg s$^{-1}$ cm$^{-2}$) & short-term / long-term &         & \\
%\tableline
24 & 03:37:58.5   & -35:19:42    & 16.7    & 379$\pm$27   & 24.9  & 24$\pm$2      & 48$\pm 2$              &  61.6 & 75$\pm 7$  & n/y & y & - \\
 
25 & 03:37:34.8   & -35:24:04    & 17.3    & 64$\pm$19    & 5.3   & 4.0$\pm$1.2   &  8.3$\pm 0.4$	   &  10.6 &  -	        & n/$\cdots$ & - & - \\
	
26 & 03:38:43.1   & -35:33:10    & 17.3    & 73$\pm$21    & 5.6   & 4.3$\pm$1.2   &  9.0$\pm 0.5$	   &  11.6 &  -	        & n/$\cdots$ & y & - \\
	
29 & 03:39:43.0   & -35:24:11    & 24.3    & 143$\pm$27   & 8.6   & 10$\pm$2      & 21$\pm 1$	           &  26.7 & 62$\pm 8$	& n/y & y & - \\
	
30 & 03:38:31.0   & -35:14:17    & 26.4    & 586$\pm$38   & 26.5  & 38$\pm$2      & 79$\pm 4$              & 102.1 & -          & 99\%/$\cdots$ & y & FCCB 1263\\
	
31 & 03:38:17.1   & -35:15:57    & 24.5    & 124$\pm$31   & 7.2   & 8$\pm$2       & 16.4$\pm 0.8$	   & 21.1  & -	        & n/$\cdots$ & y & - \\
	
32 & 03:38:25.7   & -35:25:27    & 24.5    & 145$\pm$37   & 6.2   & 8.5$\pm$2.2   & 17.7$\pm 0.9$	   & 22.8  & 27$\pm 7$	& n/n & y & - \\
	
33 & 03:37:36.5   & -35:33:37    & 24.2    & 921$\pm$40   & 41.4  & 60$\pm$3      & 125$\pm 6$             & 161.5 & -          & n/$\cdots$ & y & - \\
	
34 & 03:37:39.4   & -35:34:17    & 21.9    & 426$\pm$31   & 24.2  & 28$\pm$2      & 58$\pm 3$            &  74.5 & -          & n/$\cdots$ & y & - \\
	
35 & 03:37:50.1   & -35:21:27    & 34.6    & 139$\pm$37   & 5.8   & 8.4$\pm$2.3   & 17.6$\pm 0.9$	   &  22.6 & -	        & n/$\cdots$ & y & - \\
	
36 & 03:39:27.8   & -35:18:52    & 34.6    & 129$\pm$35   & 5.6   & 8.6$\pm$2.3   & 18.0$\pm 0.9$	   &  23.2 & 46$\pm 8$	& n/y & y & - \\
	
37 & 03:37:36.3   & -35:14:47    & 49.0    & 153$\pm$44   & 5.1   & 11$\pm$3      & 24$\pm 1$	           &  30.7 & 25$\pm 7$  & n/n & - & - \\
	
38 & 03:37:33.4   & -35:17:17    & 49.0    & 190$\pm$49   & 6.0   & 13$\pm$3      & 27$\pm 2$	           &  35.2 & -	        & n/$\cdots$ & - & - \\
	
39 & 03:37:29.0   & -35:41:11    & 69.3    & 580$\pm$119  & 14.1  & 67$\pm$14     & 140$\pm 7$            & 180.1 & 232$\pm 12$ & n/$\cdots$ & y & - \\
	
40 & 03:38:14.7   & -35:17:57    & 69.3    & 244$\pm$67   & 5.2   & 15$\pm$4      & 31$\pm 2$	           &  40.3 & -	        & n/$\cdots$ & - & - \\
	
41 & 03:38:36.7   & -35:41:32    & 62.9    & 366$\pm$61   & 8.9   & 24$\pm$4      & 51$\pm 3$	           &  65.8 & 57$\pm 7$  & n/n & - & - \\
	
42 & 03:39:12.3   & -35:26:12    & 98.0    & 312$\pm$91   & 4.7   & 19$\pm$5      & 39$\pm 2$	           &  50.7 & -	        & n/$\cdots$ & - & - \\
	
43 & 03:38:12.2   & -35:30:02    & 98.0    & 365$\pm$101  & 5.0   & 22$\pm$6      & 46$\pm 2$              &  59.1 & 60$\pm 14$ & n/n & - & - \\

44 & 03:38:05.1   & -35:19:05    & 12.2    & 42$\pm$13    & 4.6   & 2.5$\pm$0.8   & 5.3$\pm 0.3$	   &   6.8 & -	        & n/$\cdots$ & - & - \\

45 & 03:37:59.9   & -35:23:30    & 17.3    & 56$\pm$18    & 4.4   & 3.3$\pm$1.0   & 6.8$\pm 0.4$	   &   8.8 & -	        & n/$\cdots$ & - & - \\

46 & 03:37:59.2   & -35:11:37    & 49.0    & 134$\pm$42   & 4.4   & 9.9$\pm$3.1   & 20$\pm 1$	           &  26.6 & 28$\pm 8$	& n/n & - & - \\
\tableline
\end{tabular}
\tablenotetext{a}{\small Short-term = probability of the Kolmogorov-Smirnov test: `n' means no variability detected at the 95\% level.\\
		  Long-term = difference between PSPC and HRI counts: `y' means variable source ($>3\sigma$), `n' means no variability detected, `$\cdots$' means no available data.}
\tablenotetext{b}{\small CGF=\cite{Hilk99a}; FCCB=\cite{Smith96}}
\end{table}
\end{landscape}
\topmargin=0in
%#######################################################################

We tried to determine which X-ray sources belong to the Fornax cluster both by visual inspection of the photographic DSS plates and by comparing their positions with those of published catalogs of the Fornax region.
Sources No.2 and 11 are likely to belong to NGC 1399 because of their position within the optical effective radius ($r_{eff}=44.7"$, \citealp{gou94}) while sources No.16 and 32 appear to be coincident with the position of two globular clusters \citep{Kiss99}. 
Two sources are clearly associated with background objects: source No.4 corresponds to a background galaxy located at $z=0.1126$ and represents the X-ray counterpart of the strongest component of the radio source PKS 0336-35 \citep{car83}; source No.30 is coincident with two optical background galaxies classified as interacting by \cite{Ferg89} and clearly visible on the DSS plates.
For the remaining sources there is no clear indication of whether they belong to the Fornax cluster or not.  

Sources No.8, 12, 19, 21, 30 are variable at the 95\% confidence level, according to our KS tests. Source No.30
is associated with a short X-ray burst occurred in the background galaxy between January and August 1996. In fact the source is not detected in both the August 1991 PSPC (RP600043n00) and the February 1993 HRI observations (RH600256n00). It is present, instead, in the last two HRI observations (RH600831n00 and RH600831a01) with a decrease in the observed count rate of a factor 8 between the two. 

We must notice that our background level was also variable so that the presence of temporal variability was  clearly detect only for the strongest sources ($\mathit{f}_{X (0.5-2.0~keV)}\geq 3\times 10^{-14}$ erg s$^{-1}$ cm$^{-2}$).  However, since the background subtraction was performed locally (i.e. extracting the background in a region surrounding the source, see \citealp{Dam97a}), this variability does not affect the source counts reported in Table \ref{source_tab}. To further check the reliability of the wavelets algorithm, we extracted the counts manually for each source using a local annulus to measure the background, finding consistent results (within 1$\sigma$) with those derived by the wavelets technique.

We cross-checked our source list with the list of sources detected on the PSPC image by the GALPIPE project (see $\S$\ref{model}). Several sources detected in the HRI field were bright enough and sufficiently isolated to allow a comparison of the PSPC flux with the HRI results and to check for long-term variability. We converted the PSPC count rates to fluxes in the 0.5-2.0 keV band (1 cps=6.397$\times 10^{-12}$ ergs cm$^{-2}$ s$^{-1}$), 
assuming the same power law spectral model  used for converting the HRI counts (see below). The PSPC fluxes and long-term variability (i.e. when PSPC and HRI flux estimates differ more than 3$\sigma$) are included in Table \ref{source_tab}.

This comparison also confirmed that sources No.39 and 46, both near the very edge of the HRI FOV, are not an artifact of the poor S/N ratio or the steep emission gradient because both are present in the PSPC data as well and detected with high significance.

%#######################################################################
\begin{figure*}[t!]
\centerline{
\subfigure[]{\psfig{figure=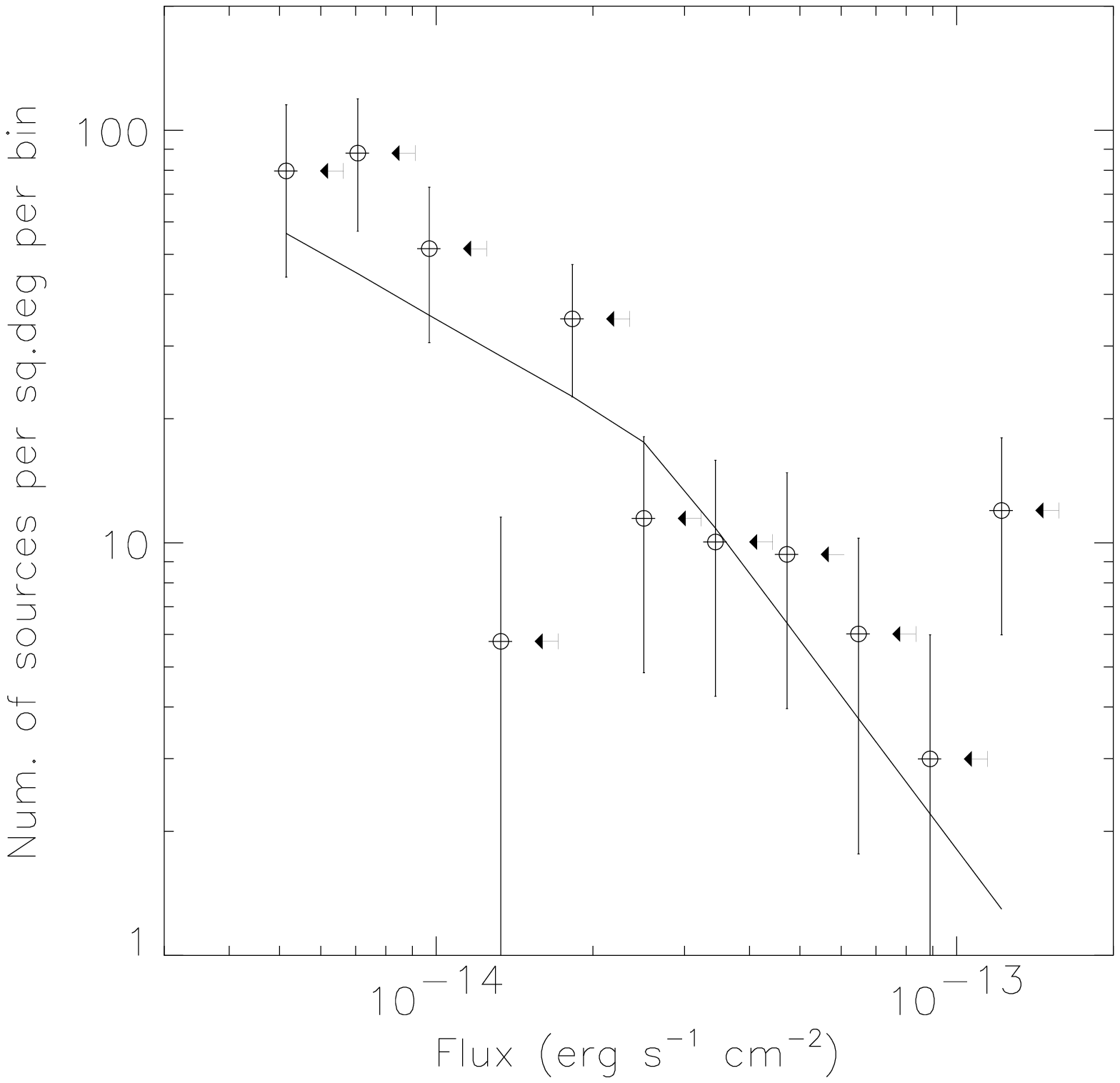,angle=0,width=0.47\textwidth}}
\subfigure[]{\psfig{figure=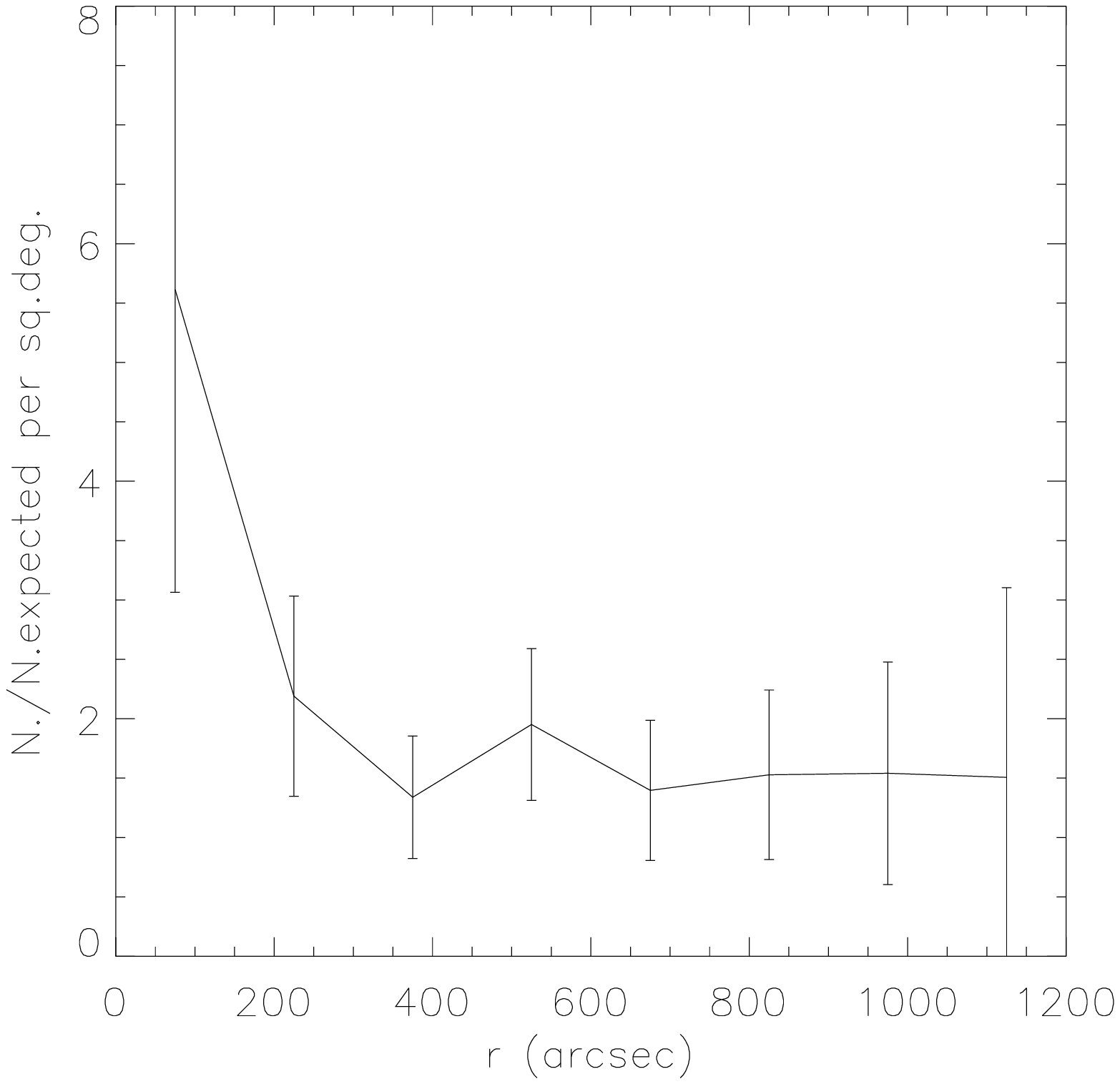,angle=0,width=0.45\textwidth}}}
\caption{{\bf (a)} Measured vs expected source counts. Measured counts, corrected for sensitivity variations across the detector, are represented by empty circles, while the continuous line is the expected distribution from \cite{Has93}. Horizontal error bars take into account uncertainties in the power law slope. Arrows represent upper limits obtained fixing the absorption to $4.5\times 10^{20}$ cm$^{-2}$ \citep{rang95}.{\bf~(b)} Radial excess of sources over the expected number from \cite{Has93}. The excess is concentrated around NGC 1399, thus suggesting to be due to sources associated with the galaxy.}
\label{src_flux}
\end{figure*}
%#######################################################################

We compared our source flux distribution to the expected background counts obtained from a deep survey of the Lockmann hole \citep{Has93}. We used the sensitivity map generated by the wavelet algorithm to weight each source for the effective detector area above the corresponding limiting flux. In this way we are able to normalize the measured counts at each flux to the total surveyed area.
We computed the flux of each source assuming a power law spectrum with photon index $1.96\pm0.11$ ($\mathit{f}_{HRI (0.5-2.0~keV)}^{pow.law}$ in Table \ref{source_tab}), as found by Hasinger and collaborators for faint sources. This gives a conversion factor of 1 cnt s$^{-1}$=$2.087\times 10^{-11}$ ergs s$^{-1}$ cm$^{-2}$ in the 0.5-2.0 keV band (the same used by Hasinger). We also tried a Raymond-Smith spectrum with $kT=0.52$ and abundance $\sim 0.2$ solar, following the results of \cite{Kim95} in their study of pointlike sources near NGC 507 and obtaining an almost identical result.

The flux distribution is shown in figure \ref{src_flux} where our differential counts (empty circles) are compared to the Hasinger ones (continuous line). The horizontal error bars take into account uncertainties in the the power law slope. Arrows represent the upper limit obtained fixing the absorption to the upper limit of $4.5\times 10^{20}$ cm$^{-2}$ ($\mathit{f}_{X (0.5-2.0~keV)}^{max~abs.}$ in Table \ref{source_tab}), allowed from the spectral analysis performed by RFFJ. Our counts are higher than the Hasinger estimate, giving a total of $312\pm 47$ sources vs $230\pm 15$ expected counts, but still compatible within 2$\sigma$. To take into account absorption uncertainties we estimated a lower limit on the number of expected sources adopting the higher absorption value found by RFFJ in the NGC 1399 field. In this case the number of expected counts decreases to $184\pm 14$ leading to an upper limit of the observed excess of $\sim 2.6\sigma$. 

The spatial distribution of detected sources is not uniform. In Figure \ref{src_flux} we show the source excess in circular annuli centered on NGC 1399. We find that the excess is peaked on the dominant galaxy, suggesting that the inner sources are likely to be associated with NGC 1399 rather than being background objects. In fact, excluding the central 300'' (28 kpc), the measured counts drop to $244\pm37$, in agreement within less than $1\sigma$ from the expected value. 
The X-ray luminosity of the sources within the central 300" range from $1.7\times 10^{38}$ erg s$^{-1}$ to $1.4\times 10^{39}$ erg s$^{-1}$, if they are in NGC 1399. Thus all of them have luminosities in excess of $1.3\times 10^{38}$ erg s$^{-1}$, the Eddinghton luminosity for a solar mass accreting object, suggesting that they may have massive Black Hole companions if they are accretion binaries.

\subsection{The {\it Chandra} data}
\label{Chandra}
While this work was in progress the {\it Chandra} data on NGC 1399 became public. 
A full data reduction of the {\it Chandra} data, which is however already in progress \citep{Ang01,Loew01}, is beyond the aim of the present work.
However we decided to compare the ROSAT data with the {\it Chandra} image to see if our conclusions are supported by the improved {\it Chandra} resolution and sensitivity.

\begin{figure}[t!]
\centerline{\psfig{figure=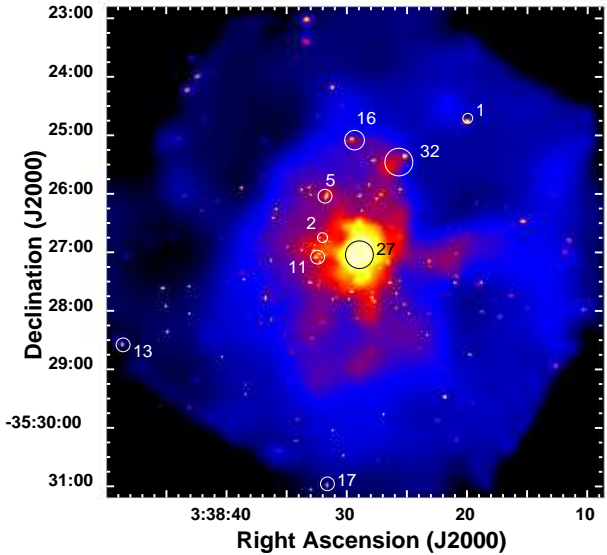,angle=0,width=0.45\textheight}}
\caption{Adaptively smoothed {\it Chandra} ACIS-S image of NGC 1399 in the energy range 0.3-10 keV. Colors from black to white represent count rates from 0.021 cnts s$^{-1}$ arcmin$^{-2}$ to 7.7 cnts s$^{-1}$ arcmin$^{-2}$. Circles represent the position of sources detected by the wavelets algorithm in the HRI image.}
\label{Chandra_csmooth}
\end{figure}

We used the longest {\it Chandra} observation of 60 ks (Obs. ID 319, 55.942 ks live time), performed on 2000 January 18 with the ACIS-S detector.
We produced a 0.5 arcsec/pixel image in the energy band 0.3-10 keV, from the ACIS-S3 chip, which covers the central $8\times 8$ arcmin of NGC 1399.
The image was then adaptively smoothed with the {\it csmooth} algorithm included in the CIAO package. The final image is shown in Figure \ref{Chandra_csmooth}.
No exposure correction was applied because the {\it csmooth} algorithm works better on the uncorrected data. However the exposure map is quite flat and we further checked that any feature present in the exposure map does not affect our conclusions. 

%#############################################
%#############################################

Figure \ref{Chandra_csmooth} shows a large number of pointlike sources present in the NGC 1399 field, most of which were unresolved in the HRI image. We marked with circles the sources detected in the HRI field (see $\S$ \ref{sources}).
The {\it Chandra} data show that sources 2, 11 and 5, identified as single sources in the HRI image, are instead multiple sources. 

The filamentary structures found in the HRI image (Figure \ref{centbox}) are present in the {\it Chandra} image as well. The `arm' protruding on the Western side and the `voids' in the X-ray emission are clearly seen. The hypothesis that the filamentary structures may be due to the presence of a large number of pointlike sources unresolved in the HRI image, is ruled out by {\it Chandra} data. In fact the {\it csmooth} algorithm clearly separates point sources from the extended structures (Figure \ref{Chandra_csmooth}) and shows that the two components are not spatially correlated.

\begin{figure}[t!]
\centerline{\psfig{figure=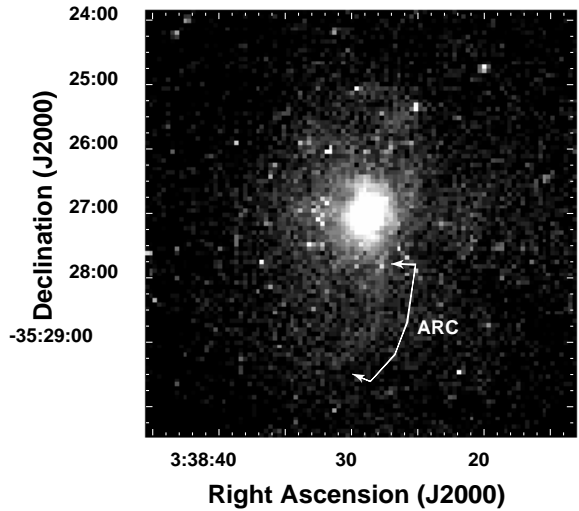,angle=0,width=0.5\textheight}}
\caption{Raw 4 arcsec/pixel {\it Chandra} image of NGC 1399. The long arc following the Western edge of the radio lobe is visible South of NGC 1399.}
\label{chandra_arc}
\end{figure}

The correspondence between the halo features and the radio jets/lobes, discussed in $\S$ \ref{Xradio}, are confirmed. Moreover the {\it Chandra} data revealed the presence of a long arc extending South that follows closely the Western edge of the radio lobe. The arc is better visible on the raw {\it Chandra} image shown in Figure \ref{chandra_arc}. This feature has a low statistical significance ($2\sigma$) but the correlation with the radio lobe argues in favor of a real structure.

%#############################################
%#############################################

\section{DISCUSSION}
\label{discussion}
The study of galactic and cluster gaseous halos has often been based on the assumption of homogeneity and spherical symmetry. The deep HRI image of the NGC 1399 halo has shown instead significant departures from these simplified assumptions. Here we briefly summarize the result of our analysis, and then discuss their physical implications.

We model the X-ray brightness distribution of NGC 1399 with three distinct components: (1) a central component (dominating for $r<50"$), coincident with the optical galaxy; (2) a `galactic' component ($50"<r\leq 400"$) displaced South-West of NGC 1399 and (3) an ellipsoidal `cluster' component ($r>400"$) centered North-East of the optical galaxy. The cooling times of the central and galactic components are smaller than the Hubble time, suggesting the presence of a central cooling flow. The binding mass estimates, obtained assuming hydrostatic equilibrium, revealed that the dynamics of the central component are dominated by stellar matter while, at radii larger than 2 arcmin there is a large dark matter halo with two distinct components dominating respectively on galactic and cluster scales.  
 
We also found significant filamentary structures and cavities in the X-ray emission. Some of these structures are correlated with the morphology and the position of the radio jets and lobes of NGC 1399, suggesting interactions between the radio emitting plasma and the hot gas.

For what concerns NGC 1404 the surface brightness distribution is well represented by a single Beta profile out to 90''. At larger radii the effect of ram pressure stripping becomes evident as a sharp brightness cutoff in the North-West
sector and an elongated tail on the opposite side of the galaxy. The cooling times derived by the deprojected density profile are shorter than the Hubble time inside the NGC 1404 halo. The mass profile shows no evidence of multiple components.
 
Finally we used a wavelets based algorithm to detect 46 sources in the HRI field, 8 of which show significant time variability. Their distribution is consistent with the background source distribution at the $3\sigma$ level.  However there is excess over background counts centered on NGC 1399. The presence of a large number of pointlike sources near the galaxy center is further confirmed by {\it Chandra} data.

\subsection{Origin of the Inner Peak of the X-ray Surface Brightness ($r<50"$)}
\label{inner}

The center of the gaseous halo is occupied by a strong emission peak (see Figure \ref{csmooth}a) centered on the optical galaxy. Such emission is often found in cooling flow galaxies (\citealp{can83,Kim95}; see also \citealp{fab91}) because of the higher density of cooler hot gas in the galactic center. RFFJ revealed the presence of a cooler region within the central 1.5 arcmin (14 kpc); this is compatible with the extent of the region dominated by the X-ray peak, as shown by the radial profile in Figure \ref{model_prof}. Moreover both the central and galactic components have cooling times below the Hubble time (assumed to be $\sim 10^{10}$ years) within respectively 250'' (23 kpc) and 350'' (32 kpc), so that significant cooling must have taken place.
The cooling time of the cluster component, instead, is compatible with the Hubble time, suggesting that the cooling flow is mainly of galactic origin.

The alternative hypothesis that the central peak can be due to the presence of a nuclear source \citep{jones97} is clearly ruled out by our data, because the observed profile (Figure \ref{profile}) is wider than the HRI PRF even taking into account the PRF uncertainties. The difference between the optical (with slope $\alpha\sim -1.6$, \citealp{Hilk99b}) and X-ray profiles ($\alpha\sim -2$ for the central component, $\S$ \ref{NGC 1399}) seen in Figure \ref{pies} also suggests that is unlikely that the central peak could be due to stellar sources. \cite{allen00} analyzing ASCA data, found that the addition of a power-law component to their NGC 1399 spectral model significantly improves the quality of the fit. They notice that the luminosity of the power-law component is compatible with the contribution expected from a population of unresolved hard X-ray binaries. However, the HRI count rate that would be produced by this power-law component falls more than an order of magnitude below the inner component
count rate. We can thus exclude that this hard discrete population dominates the central component emission.

The gravitational mass estimate derived from the hydrostatic equilibrium (equation \ref{mass_eq}) shows that the hot gas within 100" follows closely the stellar mass profile. This result is independent on the assumed temperature profile (Figure \ref{mass_prof}) being mostly determined by the hot gas density profile. These X-ray data are thus in excellent agreement with the optical mass determinations of \cite{saglia00} who find no evidence of dark matter in the galaxy core.

The origin of the central density enhancement may be due to hot gas inflow within the inner 1 arcmin. The deprojection technique described in $\S$ \ref{dens_par} allowed to derive the cumulative gas mass profile of the central component (Figure \ref{mass_prof}). According to the  mass deposition rate measured by RFFJ, which is in good agreement with our data, we expect to find $\sim 2.5\times 10^{9} {\rm ~M}_\odot$ of deposited gas within the central 60'' ($\sim 5$ kpc) over $10^{10}$ yr, while we only observe $\sim 2\times 10^8 {\rm ~M}_\odot$ in the hot phase. Taking into account a possible larger than line-of-sight central absorption, the observed mass would increase of $\sim 30\%$, but the deposition rate would also increase (see RFFJ), so that the result remains almost unchanged. 
This difference suggests that $\sim 90\%$ of the hot gaseous mass that reached the inner regions of NGC 1399 has cooled enough to become invisible in X-rays. However RFFJ argue that the missing mass is not seen in any other phase.
The situation gets even worse considering that we are ignoring stellar mass losses which must be considerable since the metallicity of the ISM in the galactic center is consistent with stellar values \citep{Buo99,Matsu00}. If the age of the cooling flow is lower than $10^{10}$ yr, this disagreement would be reduced. Equating the observed hot gas mass to the amount expected from the cooling flow, we estimate a lower limit of $10^9$ years to the  age of the cooling flow. 
However, we must notice that the latest studies of cooling flows in clusters and galaxies with Newton-XMM are showing that the classical cooling flow model is not able to reproduce the observed temperature distributions \citep{Pet01,Kaa01,Tam01}. In this case no conclusion can be drawn from our data until the cooling flow physics will be better understood.

\subsection{Dynamical Status of the Halo}
\subsubsection{Gravitational Effects}
The X-ray brightness profiles (Figure \ref{pies}) shows that past 1 arcmin the hot gas distribution is more extended than the stellar distribution. If the gas is in hydrostatic equilibrium, additional matter is required to explain this broader distribution. The change in slope of the X-ray brightness profile for $r>400"$ further requires the presence of two components that we identified with the `galactic' and `cluster' halo. This complex distribution was previously found by \cite{ikebe96} using the ASCA data, and explained in terms of a different concentration of dark matter on galactic and cluster scales \citep{ikebe96,Maki01}. Our data further show that this galactic component is not correlated with stellar matter, which is even more concentrated (Figure \ref{mass_prof}b) and dominates the dynamics of the galactic core.

The kinematics inferred from the GCs and PNe (see Figure 6 of \citealp{Kiss99})  indicate that for $1'<r<16'$ (5 kpc$<r<$90 kpc) from the galaxy center, the velocity dispersion increases with radius. This result is consistent with our measurements and suggests either the presence of a significant amount of dark matter associated with the outer radii of NGC 1399 or equivalently that the outer galactic dynamics is dominated by the cluster potential.

We must notice however that while on galactic scales the mass profiles may be explained by the presence of dark matter, the deviations from symmetry, i.e. the offset between the stellar component (consistent with the central component) and the galactic dark halo, are hardly explained in the galaxy formation framework. For this reason in next sections we take into account possible departures from equilibrium.

\subsubsection{Galaxy-Galaxy Interactions}   
An alternative scenario has recently been advanced by Napolitano, Arnaboldi \& Capaccioli (2001, in preparation), that could explain our results by means of galaxy interactions.
 Their calculations show that the energy injection due to galaxy-galaxy encounters can account for the outer galaxy dynamics as seen in PNe and GCs, and is in agreement with the observed X-ray temperature profile. In this case there would be no need for additional dark matter in the `galactic' halo. 
Support for the tidal interaction picture comes from \cite{Kiss99} who showed that the GC system of NGC 1399 is 10 times more abundant than those of neighbors galaxies, while their properties are very similar. In the hypothesis that all galaxies initially had the same number of GCs per unit luminosity, the excess around NGC 1399 may be compatible with tidal stripping of GCs from nearby Fornax galaxies.  

The effects of tidal interactions on the gaseous halos of elliptical galaxies have been explored by \citet[ hereafter DRC]{D'Ercole00}. They found that the interaction strongly affects the gas flow, altering the ISM density and thus increasing the cooling speed and luminosity. This mechanism is enhanced when energy injection from SNe Ia is included and the gas remains turbulent independently from the epoch of the encounter. In this process the surface brightness is scarcely affected and the global isophotes remain circular, although locally distorted. 
The DRC model predicts that the gas inflow produces steeper profiles in the central regions of the galaxy in the moment of maximum luminosity, indicating that the tidal perturbations may contribute to the formation of central brightness peaks such as the one seen in the NGC 1399 core. The steep slope of the galactic component (Table \ref{fit_tab}) too may be due to tidal gas stripping because the galaxy undergoes significative mass losses (up to 50\%) in the outer regions.
However, we must notice that in the case of NGC 1399, the effect of tidal stripping may be smaller than what expected from the DCR models because PSPC data \citep{jones97} suggest a lower SN rate than assumed by D'Ercole and collaborators.

For these models to be consistent with the X-ray observations, tidal interaction must strongly affect the halo density distribution: the increase in the M/L ratio past 2 arcmin (Figure \ref{mass_prof}) is mainly determined by the shallower slope of the gas density profile in the outer galactic regions, since the isothermal model still exhibits the multi-component behavior ($\S$ \ref{dens_par}). Thus a simple temperature increase would not produce the observed agreement with the optical data. Such strong profile alterations are not seen in the DRC simulations. Conversely \cite{bar00} found that encounters between galaxies of different masses can produce both radial profiles with a central peak and a `shoulder' similar to our multi-component structure and azimuthal asymmetries as those seen in the NGC 1399 galactic halo. However his simulations also show that the stellar profile follows the gas profile, which is not seen in our data. 

The X-ray surface brightness analysis ($\S$ \ref{brightness}) revealed the presence of complex structures in the galactic halo. We have shown in $\S$ \ref{model} that these brightness fluctuations are not due to artifacts of the smoothing algorithm, because they are present in the residuals of the bidimensional model, and have high statistical significance. Due to the weak dependence of the HRI spectral response on the temperature around 1 keV (see \citealp{clar97}), they must be due to local density fluctuations of the hot gas. While some of these features can be due to the interaction between the radio jets and the surrounding ISM ($\S$ \ref{Xradio}), most of them are not related to any evident radio or optical feature and are spread all around the galaxy, suggesting a `disturbed' nature of the whole halo.

It is possible that these structures are the signature of interaction between the NGC 1399 halo and other Fornax galaxies. In fact DRC predict significative density fluctuations in the hot gas and the formation of filamentary structures near the galaxy center as a consequence of galaxy-galaxy encounters. It's not easy to determine if these perturbations are due to the nearby NGC 1404 or to more distant members because, according to DRC, the perturbing effects lasts several Gyrs. We searched for possible signatures of interactions between NGC 1399 and NGC 1404 looking for excess emission in the region between the two galaxies. The slight excess we found is only significant at the 2$\sigma$ level, so that no significant conclusion can be drawn from our data.

\subsubsection{Ram Pressure}
A different scenario that could explain the cluster and galactic halo displacement is one in which NGC 1399 is moving in the Fornax cluster potential, thus experiencing ram pressure from the cluster hot gas. The radial profiles in Figure \ref{pies} show a high degree of asymmetry with respect to the optical galaxy. The 2D models indicate that the galactic component is displaced $\sim 1'$ (5 kpc) South-West with respect to the X-ray peak (and thus optical galaxy), while the cluster component seems located $\sim 5.6$ arcmin (31 kpc) North-East of NGC 1399. \cite{jones97} argue that this displacement is compatible with the hypothesis that the galaxy is slowly moving (40 km s$^{-1}$) South-East in a larger potential. This would explain the displacement of the galaxy with respect to the cluster component but not the one of the galactic halo. In the ram pressure scenario, the displacement and large scale asymmetry of the `galactic' halo may suggest that NGC 1399 is not sitting at the very center of the cluster potential and is slowly moving North-East toward the `cluster' component.

The effect of ram stripping has already been observed in nearby galaxies. For instance \cite{irw96} observed the presence of a very irregular halo in the Virgo galaxy NGC 4472. They found a steeper gradient in the X-ray emission in the direction of the cluster center associated with a temperature enhancement. They calculate that ram stripping is able to explain the observed halo features if the galaxy is moving at $\sim 1300$ km s$^{-1}$ in the cluster potential. Their ROSAT HRI images also revealed filamentary structures and holes in the galactic halo, very similar to the features that we observed in the adaptively smoothed image of the inner halo (Figure \ref{centbox}). 
Additional evidence that galaxies moving in the cluster environment can produce filaments of cooler gas behind them were found by \cite{Dav94} or \cite{fabian01}. 

The accurate simulation of dynamic stripping in cluster and group elliptical galaxies performed by \cite{toni01}, show that the effect of dynamic stripping depends on the galaxy orbit and the duration of the subsonic and supersonic phases which determine the balance within gas losses and accretion onto the galaxy. In general their simulations reveal the formation of complex and elongated tails behind the galaxy (similar to those seen in NGC 4472) and strong decentering of X-ray isophotes due to supersonic stripping.
Such irregular structures are not seen in the NGC 1399 galactic halo which seems instead rather circular ($\epsilon_{Galactic}=0.02$).

A very rough estimate of the velocity of NGC 1399 can be obtained supposing that the displacement of the optical galaxy, coincident with the X-ray centroid, with respect to the galactic halo is due to the deceleration of the halo caused by ram pressure from the cluster component. In the hypothesis that the motion is highly subsonic, we can treat the different components as homogeneous spheres and calculate the initial velocity required to produce the observed displacement after a time $t=10^9$ yr due to the ram pressure of the cluster halo $P\sim \rho_{cluster} v^2$. Using the parameters derived form the bidimensional model ($\S$ \ref{model}): $\rho_{galaxy}/\rho_{cluster}\sim 4$, $R_{galaxy}\sim 4'$ and a displacement $\Delta S\sim \frac{1}{2} R$, we obtain an initial velocity of $\sim 10$ km s$^{-1}$, consistent with the initial assumption of subsonic motion.
Even considering a more recent interaction where $t=10^8$ the galaxy motion remains highly subsonic ($\sim 100$ km s$^{-1}$). 
This result is in agreement with the observations, showing that the halo morphology is only weakly distorted. 
In conclusion, even though extreme projection effects may hide stronger  stripping features, ram pressure may be causing the galaxy halo displacement but seems unlikely to be responsible for the observed inner halo structure of NGC 1399.

\subsubsection{Cluster Merging effects}
For what concerns the elongated cluster halo component, in many cases irregular features in the cluster halos seems associated with merger events (see for instance \citealp{sun00,mazz01}). Simulations of cluster merging \citep{roett96} show that elongation and twisting of the X-ray isophotes are a general result of such interactions. In the case of the Fornax cluster \cite{drink00} found evidence of a subcluster 3 degrees South-West of NGC 1399 centered on NGC 1316 (Fornax A). They claim that the two components are at the early stage of merging; in this case merging is unlikely to be the cause of the elongation of the cluster X-ray isophotes. To further study the dynamical status of the cluster halo we would need temperature maps of the whole cluster, as can be obtained with {\it Chandra} or XMM.

It is worth noticing that the X-ray centroid of the cluster component is displaced with respect to the optical distribution of Fornax cluster galaxies, which tend to concentrate South-West of NGC 1399 (see Figure 1 from \citealp{Hilk99a} and Figure 3 from \citealp{drink00}). If the merging picture is confirmed, this may imply that the galaxies in the two subclusters are infalling in the global potential well faster than the associated gaseous halos.

\cite{Kim95} compared three X-ray dominant galaxies in poor groups: NGC 1399, NGC 507 and NGC 5044. They show that even if these galaxies have similar properties (temperatures, cooling core) their X-ray brightness profiles are very different. This suggests that the shape of the potential and the mass deposition mechanism may vary from galaxy to galaxy. If environmental effects (tidal interactions, ram pressure stripping) are important factors in determining the NGC 1399 halo structure, they must be considered as well in order to explain the observed differences.

\subsection{The NGC 1404 halo}
In contrast with NGC 1399, NGC 1404 possesses a very regular halo, well represented by a Beta profile. The X-ray brightness profile falls as $\sim r^{-2.08\pm 0.03}$, in good agreement with the optical brightness profile $\Sigma\propto r^{1.9\pm 0.1}$ \citep{Forb98}. This result suggests that the the gas distribution of the hot halo of NGC 1404 is produced by the same gravitational potential that is binding the stellar population, out to $r\sim 80"$ ($\sim 7$ kpc) from the galaxy center (Figure \ref{1404profile}b).
This result is in agreement with \cite{Matsu01} who found that luminosities and temperatures of X-ray compact galaxies are well explained by kinematical heating of gas supplied by stellar mass loss. However she notice that NGC 1404 is significantly brighter than other compact ellipticals and explains this result in term of interaction with the ISM.

Indeed at radii larger than 80'' the influence of dynamic stripping is evident from  the presence of an elongated tail on the South-East side of the galaxy and the steep surface brightness gradient on the opposite side (Figure \ref{csmooth} and \ref{1404profile}b). However the regularity of the brightness profile and its similarities to the optical profile within 90" ($\sim 8$ kpc) implies that either the stripping efficiency is low or that it affects significantly only the outer galactic regions.
It is worth noticing that our brightness profile is shallower than expected from simulations of dynamically stripped galaxies suggesting that the models do not properly describe the influence of ram stripping on the inner galactic halo (see  \citealp{toni01}).

The gas and total mass profiles derived in $\S$ \ref{dens_par_1404} are in the same range of those of the central component of the NGC 1399 halo, within the inner 80''. This is not surprising since we have seen that in NGC 1399 the central component seems to be in equilibrium with the stellar population, as it is the case for the NGC 1404 halo. This result further supports the scenario in which the difference between X-ray compact and extended galaxies lies in the presence of additional components (i.e. galactic and cluster components) related to the central position of the dominant galaxies inside the cluster potential well (e.g. \citealp{Matsu01}).

\subsection{X-ray--Radio Interactions}

Signs of interaction between radio jet/lobes and the surrounding gaseous medium have been reported in powerful radio galaxies (see for instance \citealp{car94} for Cygnus A;  \citealp{bor93,McNam00,Fab00} for evidence of such interaction in clusters hosting powerful radio sources).
NGC 1399 is a faint radio galaxy whose radio lobes are entirely contained within the optical galaxy. Nevertheless the structures seen in Figure \ref{X-radio} suggest that the same mechanisms at work in more powerful sources may be responsible for some of the galactic halo structures. 
The residual image in Figure \ref{X-radio_nuclear} shows excess emission on both sides of the nucleus and a lack of X-ray emission along the radio jets in the inner 15" ($\sim 1.4$ kpc), as if the radio jets are displacing the ambient gas. For $r\sim 25"$ ($\sim 2.3$ kpc) more X-ray emission is seen in the North and South part of the galaxy, aligned with the direction of the radio jets. 

\cite{clar97} have studied, by means of hydrodynamical simulations, how these features are produced during the propagation of the radio jet trough the ICM. They show that, for a $\sim 1$ keV gaseous halo and a detector response similar to the ROSAT HRI (see their Figure 8), we expect to see excess emission located on the side of the jets, due to the shocked gas near the tail of the radio lobes, as the one seen in  Figure \ref{X-radio} (region {\bf A}) and in Figure \ref{X-radio_nuclear}. 

In coincidence with the radio lobes, X-ray enhancements or deficits are expected depending on the thickness of the shocked gas $\Delta r$ with respect to the lobe radius $r$. Near the edge of the radio lobe, where $r\geq 3.4\Delta r$, the shocked gas density is high enough to produce a brightness enhancement. On the contrary, in the tail of the lobes the compressed gas had time to expand, so that $r<3.4\Delta r$ and the X-ray brightness is diminished, producing a `cavity'. This picture is very similar to what we see in our data.
Features {\bf B} and {\bf C} in Figure \ref{X-radio} may suggest respectively the presence of a cavity produced by the Southern lobe in the ICM and the excess emission associated with the edge of the lobe. The alternative hypothesis, that the excess is due to the presence of a bow shock at the leading end of the lobe, seems unlikely both because Carilli and collaborators showed that the bow shock is hardly seen at energies $< 4$ keV and because the signatures of bow shocks are usually sharper than the one seen in our data, showing up as compact bright spots or arcs.  The {\it Chandra} image also revealed the presence of a long arc South of NGC 1399. The steeper gradient of the radio emission on the Western side of the Southern lobe, with respect of the Eastern side, suggests that this feature may represent a layer of hot gas compressed by the radio emitting plasma.

Signatures of interaction with the ICM do not show up so clearly in the Northern lobe, but this could be due to projection effects (see for instance \citealp{bru01,fino01}). The brightness enhancement near the West side of the lobe could be due to the thin layer of compressed
gas surrounding the lobe while the steeper radio gradient near region {\bf D} suggests that the lobe is avoiding a region of higher gas density. Indeed the crude estimate  indicates that the pressure of the bubble ($\S$ \ref{Xradio}) can be up to 4 times higher than that of the surrounding ISM. An upper limit of $k\simeq 420$ on the ratio of the electron to heavy particle energy (or alternatively a lower limit of 0.005 on the filling factor) can be derived by matching the radio pressure to the thermal pressure of the bubble (Figure \ref{press_prof}), in agreement with \cite{kbe88}.

Inverse Compton (IC) scattering of Cosmic Background photons by the radio emitting particles is often considered as an alternative process to explain the X-ray brightness enhancements near radio lobes \citep{harris79,kan95}. We have evaluated this possibility for the X-ray excess near the edge of the Southern radio lobe of NGC 1399, following \cite{Feig95}, and comparing the minimum energy field strength $B_{ME}$ with the magnetic field $B_{IC}$ required to produce a measured ratio of radio to X-ray flux. We used the radio flux for the Southern lobe measured by \cite{kbe88}, corrected for the fact that just half of the lobe contributes to the X-ray excess, and we measured the X-ray flux density from the residuals shown in Figure \ref{residuals}, assuming that all the excess over the 2D model is due to IC scattering.
We find that it is unlikely that more than a few percent of the excess near the edge of the Southern radio lobe may be due to IC scattering, even taking into account the poor knowledge of the parameters involved in the calculations. 

Pressure confinement of the radio lobes \citep{kbe88} is supported by our data (Figure \ref{press_prof}). If the inclination of the radio source is $< 45^\circ$, the radio jets remain collimated up to the radius within which the high density central component dominates ($\sim 50"$, Figure \ref{press_prof}) and diffuse in lobes as soon as they enter in the lower density environment. A larger inclination is unlikely because the thermal pressure of the cluster component (long-dashed line in Figure \ref{press_prof}) is of the same order of magnitude as the radio lobes pressure, and would probably not be able to confine the radio emitting plasma, if this was extending at greater distances.

Finally, we analized the possibility that the power-law component found in the spectral model of \cite{allen00} for NGC 1399 (see $\S$ \ref{inner}) may be due to an active nucleus. Their estimated flux (converted to our adopted distance) is twice the value expected from our upper limit on a possible nuclear source, adopting their best fit parameters. Considering that \cite{Loew01} further reduce this upper limit by an order of magnitude it appears that the bulk of the  power-law emission comes from the discrete source population revealed by the {\it Chandra} data (see Figure \ref{Chandra_csmooth}).

\subsection{Discrete sources}
The presence of a population of individual X-ray sources in E and S0 galaxies was first suggested by \cite{Trin85}. Additional evidence was then provided by ROSAT and ASCA observations (e.g. \citealp{Fab94,Kim96,Matsumo97}). These X-ray source populations are now been detected down to much lower luminosities with {\it Chandra} (e.g. \citealp{Sar00,Sar01}). The HRI data demonstrate the presence of such population, possibly of accretion binaries, also in NGC~1399. This result is confirmed by the {\it Chandra} data which reveal a large number of pointlike sources surrounding the galaxy, many of which associated with globular clusters \citep{Ang01}. This population of accretion binaries is probably responsible for the hard spectral component \citep{Buo99,Matsu00,allen00} required to fit the high energy spectrum of NGC 1399.  
As demonstrated by comparing the 5" resolution Rosat HRI image with the 10 times sharper {\it Chandra} ACIS image, subarcsecond resolution is essential for the study of X-ray source populations in galaxies.

\section{SUMMARY}
We have analized a deep ROSAT HRI observation centered on the dominant Fornax cluster galaxy NGC 1399 and including NGC 1404, in conjunction with the archival ROSAT PSPC data of the same field. We found an extended and asymmetric halo extending on cluster scales ($r>90$ Kpc). The halo profile is not consistent with a simple Beta model but suggests the presence of three different components. The combined HRI and PSPC data were fitted with a multi-component bidimensional model, consisting of: (1) a `central' component (dominating for $r<50"$) centered on the optical galaxy whose distribution follows the luminous matter profile; (2) a `galactic' component ($50"<r\leq 400"$) displaced $1'$ (5 kpc) South-West of NGC 1399 and (3) a `cluster' ellipsoidal component ($r>400"$) centered  $\sim 5.6'$ (31 kpc) North-East of the galaxy. The HRI image also revealed the existence of filamentary structures and cavities in the galactic halo, due to density fluctuations in the hot gas. The presence of these features is confirmed by a preliminary analysis of {\it Chandra} data.
The large scale surface brightness distribution can be explained if the galaxy hosts a large dark halo with different dark matter distributions on galactic and cluster scales \citep{saglia00,ikebe96}.
Alternative models that explain the GCs abundance and optical velocity dispersion profiles through tidal interactions with Fornax cluster galaxies (\citealp{Kiss99}; Napolitano, Arnaboldi \& Capaccioli 2001, in preparation) are compatible with our data only if such encounters produce a significative flattening of the outer gas distribution. Tidal interactions may also explain the presence of the density fluctuations in the galactic halo \citep{D'Ercole00}.
Ram pressure from the cluster halo (e.g. \citealp{irw96,Dav94,fabian01}) is able to account for the decentering of the optical galaxy with respect to the galactic halo if NGC 1399 is moving subsonically (10-100 km s$^{-1}$) in the cluster potential. The displacement of the center of the cluster component with respect to the nearest Fornax galaxies may suggest that the cluster is not relaxed and may be undergoing a merger event.   

We do not detect the X-ray counterpart of the nuclear radio source \citep{kbe88} and pose an upper limit of $L_X^{3\sigma}=3.9\times 10^{39}$ erg s$^{-1}$ on its luminosity in the 0.1-2.4 keV band. In agreement with Killeen and collaborators, our data support thermal confinement of the radio emitting plasma.
We found X-ray emission enhancements aligned with the radio jets that may be due to shocked gas and X-ray `holes' and enhancements coincident with the position of the radio lobes. The latter are consistent with a scenario in which the hot gas is displaced by the radio plasma pressure (e.g. \citealp{clar97}), while alternative models of Inverse Compton scattering of CMB photons fail to account for the observed excesses. 

NGC 1404 possess a very symmetric halo within 8 kpc well fitted by a simple Beta model. The agreement with the optical profile and the similarity with the 'central' component of the NGC 1399 halo suggest that the ISM distribution of NGC 1404 is produced by the same gravitational potential that is binding the stellar population. At larger radii the influence of dynamic stripping is evident from  the presence of an elongated tail on the South-East side of the galaxy and the steep surface brightness gradient on the opposite side.

We detected 43 discrete sources in our field. Their flux distribution is consistent at the $2\sigma$ level with the number expected from X-ray background counts. However, there is a significant excess `peak' on the central galaxy in the spatial distribution of these sources, suggesting a population of galaxian X-ray sources in NGC 1399. This is confirmed by the {\it Chandra} data, that reveal a large number of point-like sources associated with the optical galaxy. This sources are likely to explain the hard component found by ROSAT and ASCA spectral analysis \citep{Buo99,Matsu00,allen00}.

~\newline
We would like to thank Dan Harris for useful discussions on the radio data interpretation, F. Damiani and S. Sciortino for advice on the use of the wavelets algorithm and the GALPIPE database. Helpful comments and suggestions were provided by the referee Dr. Loewenstein and the editor Dr. Bothun.\\
We aknowledge partial support from the CXC contract NAS8-39073 and NASA grant NAG5-3584 (ADP), from the MURST PRIN-COFIN 1998-1999 (resp. G. Peres) and from the European Social Fund (F.S.E.).\\
This research has made use of the NASA/IPAC Extragalactic Database (NED) which is operated by the Jet Propulsion Laboratory, California Institute of Technology, under contract with the National Aeronautics and Space Administration.


\begin{thebibliography}{}  % (do not forget {})
\bibitem[Allen, di Matteo \& Fabian(2000)]{allen00} Allen, S. W., di Matteo, T., \& Fabian, A. C. 2000, MNRAS, 311, 493
\bibitem[Angelini, Loewenstein \& Mushotzky(2001)]{Ang01} Angelini, L., Loewenstein, M., \& Mushotzky, R. F. in press in ApJL, astro-ph/0107362  
\bibitem[Arnaboldi et al.(1994)]{Arn94} Arnaboldi, M., Freeman, K. C., Hui, X., Capaccioli, M., \& Ford, H. 1994, Messenger, 76, 40
\bibitem[Barnes(2000)]{bar00} Barnes, J. E. 2000, to appear in ``Stellar Collisions, Mergers, and Their Consequences'', M. Shara, ed., astro-ph/0010145 
\bibitem[B\"ohringer et al.(1993)]{bor93} B\"ohringer, H., Voges, W., Fabian, A.C., Edge, A.C., \& Neumann, D.M., 1993, MNRAS 264, L25
\bibitem[Brunetti et al.(2000)]{bru01} Brunetti, G., Cappi, M., Setti, G., Feretti, L., \&  Harris, D.E. 2001, A\&A, 372, 755
\bibitem[Buote(1999)]{Buo99} Buote, D. A. 1999, MNRAS, 309, 685
\bibitem[Canizares, Stewart \& Fabian(1983)]{can83} Canizares, C. R., Stewart, G. C., \& Fabian, A. C. 1983, ApJ, 272, 449
\bibitem[Carilli, Perley \& Harris(1994)]{car94} Carilli, C.L., Perley, R.A., \& Harris, D.E. 1994, MNRAS, 270, 173
\bibitem[Carter \& Malin(1983)]{car83} Carter, D., \& Malin, D. F. 1983, MNRAS, 203, 49
\bibitem[Clarke, Harris \& Carilli(1997)]{clar97} Clarke, D.A., Harris, D.E., \& Carilli, C.L. 1997, MNRAS, 284, 981 
\bibitem[Damiani et al.(1997a)]{Dam97a} Damiani, F., Maggio, A., Micela, G., Sciortino, S. 1997a, ApJ, 483, 350
\bibitem[Damiani et al.(1997b)]{Dam97b} Damiani, F., Maggio, A., Micela, G., Sciortino, S. 1997b, ApJ, 483, 370
\bibitem[David et al.(1994)]{Dav94} David, L. P., Jones, C., Forman, W., \& Daines, S. 1994, ApJ, 428, 544
\bibitem[David et al.(1996)]{Dav96} David, L. P., Harden, Jr., F. R., Kearns, K. E., \& Zombeck, M. V. 1996, The ROSAT High Resolution Imager (HRI), Calibration Report 1996 February, revised, U.S. ROSAT Science Data Center, SAO, Cambridge, MA
\bibitem[D'Ercole, Recchi \& Ciotti(2000)]{D'Ercole00} D'Ercole, A., Recchi, S., \& Ciotti, L. 2000, ApJ, 533, 799 (DRC)
\bibitem[Drinkwater, Gregg \& Colless(2000)]{drink00} Drinkwater, M. J., Gregg, M. D., \& Colless, M. 2001, ApJ, 548, L139
\bibitem[Ekers et al.(1989)]{ekers89} Ekers, R. D., et al. 1989, MNRAS, 136, 737
\bibitem[Fabbiano, Kim \& Trinchieri(1984)]{Fab94} Fabbiano, G., Kim, D.-W., \& Trinchieri, G. 1994, ApJ, 429, 94
\bibitem[Fabian, Nulsen \& Canizares(1991)]{fab91} Fabian, A. C., Nulsen, P. E. J., \& Canizares, C. R. 1991, A\&A Rev, 2, 191
\bibitem[Fabian et al.(2000)]{Fab00} Fabian, A.C., Sanders, J.S., Ettori, S., Taylor, G.B., Allen, S.W., Crawford, C.S., Iwasawa, K., Johnstone, R.M., \& Ogle, P.M. 2000, MNRAS, 318, L65
\bibitem[Fabian et al.(2001)]{fabian01} Fabian, A. C., Sanders, J. S., Ettori, S., Taylor, G. B., Allen, S. W., Crawford, C. S., Iwasawa, K., \& Johnstone, R. M. 2001, MNRAS, 321, L33
\bibitem[Fabricant, Lecar \& Gorenstein(1980)]{Fabr80} Fabricant, D., Lecar, M., \& Gorenstein, P. 1980, ApJ, 241, 552
\bibitem[Feigelson et al.(1995)]{Feig95} Feigelson, E. D., Laurent-Muehleisen, S. A., Kollgaard, R. I., \& Fomalont, E. B. 1995, ApJ, 449, L149
\bibitem[Ferguson(1989)]{Ferg89} Ferguson, H. C. 1989, AJ, 98, 367
\bibitem[Finoguenov \& Jones(2000)]{fino01} Finoguenov, A., \& Jones, C. 2001, ApJ, 547, L107
\bibitem[Forbes et al.(1998)]{Forb98} Forbes, D. A., Grillmair, C. J., Williger, G. M., Elson, R. A. W., \& Brodie, J, P. 1998, MNRAS, 293, 325
\bibitem[Graham et al.(1998)]{gra98} Graham, A. W., Colless, M. M., Busarello, G., Zaggia, S., \& Longo, G. 1998, A\&AS, 133, 325
\bibitem[Goudfrooij et al.(1994)]{gou94} Goudfrooij, P., Hansen, L., J{\o}rgensen, H. E., N{\o}rgaard-Nielsen, H. U., de Jong, T., \& van den Hoek, L. B., 1994, A\&AS, 104, 179
\bibitem[Grillmair et al.(1994)]{Grill94} Grillmair, C. J., Freeman, K. C., Bicknell, G. V., Carter, D., Couch, W. J., Sommer-Larsen, J., \& Taylor, K. 1994, ApJ, 422, L9
\bibitem[Harris \& Grindlay(1979)]{harris79} Harris, D. E., \& Grindlay, J. E. 1979, MNRAS, 188, 25
\bibitem[Harris et al.(1998)]{Harris98} Harris, D. E., Silverman, J. D., Hasinger, G., \& Lehmann, I. 1998, A\&AS, 133, 431
\bibitem[Harris(1999)]{Har99} Harris, D. E. 1999, American Astronomical Society, HEAD meeting 31, 17.12
\bibitem[Hasinger et al.(1993)]{Has93} Hasinger, G., Burg, R., Giacconi, R., Hartner, G., Schmidt, M., Trumper, J., Zamorani, G., 1993, A\&A, 275, 1
\bibitem[Hilker et al.(1999a)]{Hilk99a} Hilker, M., Kissler-Patig, M., Richtler, T., Infante, L., \& Quintana, H. 1999, A\&AS, 134, 59
\bibitem[Hilker et al.(1999b)]{Hilk99b} Hilker, M., Infante, L., \& Richtler, T. 1999, A\&AS 138, 55
\bibitem[Irwin \& Sarazin(1996)]{irw96} Irwin, J. A., \& Sarazin, C. L. 1996, ApJ, 471, 683
\bibitem[Ikebe et al.(1992)]{ikebe92} Ikebe, I., et al. 1992, ApJ, 384, L5
\bibitem[Ikebe et al.(1996)]{ikebe96} Ikebe, Y., et al. 1996, Nature, 379, 427
\bibitem[Jones et al.(1997)]{jones97} Jones, C., Stern, C., Forman, W., Breen, J., David, L., \& Tucker, W. 1997, ApJ, 482, 143
\bibitem[Kaastra et al.(2001)]{Kaa01}  Kaastra, J. S., Ferrigno, C., Tamura, T., Paerels, F. B. S., Peterson, J. R., \& Mittaz, J. P. D. 2001, A\&A, 365, L99 
\bibitem[Kaneda et al.(1995)]{kan95} Kaneda, H., Tashiro, M., Ikebe, Y., Ishisaki, Y., Kubo, H., Makshima, K., Ohashi, T., Saito, Y., Tabara, H., \& Takahashi, T. 1995, ApJ, 453, L13
\bibitem[Killeen \& Bicknell(1988)]{kill88} Killeen, N. E. B., \& Bicknell, G. V. 1988, ApJ, 325, 165
\bibitem[Killeen, Bicknell \& Ekers(1988)]{kbe88} Killeen, N. E. B., Bicknell, G. V., \& Ekers, R. D. 1988, ApJ, 325, 180
\bibitem[Kim, Fabbiano \& Trinchieri(1992)]{Kim92} Kim, D.-W., Fabbiano, G., \& Trinchieri, G. 1992, ApJS, 80, 645
\bibitem[Kim et al.(1996)]{Kim96} Kim, D.-W., Fabbiano, G., Matsumoto, H., Koyama, K., \&  Trinchieri, G. 1996, ApJ, 468, 175
\bibitem[Kim \& Fabbiano(1995)]{Kim95} Kim, D.-W., \& Fabbiano, G. 1995, ApJ, 441, 182
\bibitem[Kissler-Patig et al.(1999)]{Kiss99} Kissler-Patig, M., Grillmair, C. J., Meylan, G., Brodie, J. P., Minniti, D., \& Goudfrooij, P. 1999, AJ, 117, 1206
\bibitem[Kriss et al.(1983)]{kriss83} Kriss, G. A., Cioffi, D. F., \& Canizares, C. R. 1983, ApJ, 272, 439
\bibitem[Loewenstein et al.(2001)]{Loew01}  Loewenstein, M., Mushotzky, R. F., Angelini, L., Arnaud, K. A., \& Quataert, E. ApJ, 555, L21
\bibitem[Mackie et al.(1996)]{mack96} Mackie, G., et al. 1996, in ASP Conference Series, Vol. 101, ``Astronomical Data Analysis Software and Systems V'', G. H. Jacoby and J. Barnes eds.
\bibitem[Makishima et al.(2001)]{Maki01} Makishima, K., et al. 2001, PASJ, 53, 401
\bibitem[Matsumoto et al.(1997)]{Matsumo97} Matsumoto, H., Koyama, K., Awaki, H., Tsuru, T., Loewenstein, M.,\& Matsushita, K. 1997, ApJ, 482, 133
\bibitem[Matsushita, Ohashi \& Makishima(2000)]{Matsu00} Matsushita, K., Ohashi, T., \& Makishima, K. 2000, PASJ, 52, 685
\bibitem[Matsushita(2001)]{Matsu01} Matsushita, K. 2001, ApJ, 547, 693
\bibitem[Mazzotta et al.(2000)]{mazz01}  Mazzotta, P. ,Markevitch, M. ,Vikhlinin, A. , Forman, W. R., David, L. P., \& VanSpeybroeck L. 2001, ApJ, in press 
\bibitem[McNamara et al.(2000)]{McNam00} McNamara, B.R., Wet al. 2000, ApJ, 530, L81
\bibitem[Peterson et al.(2001)]{Pet01} Peterson, J. R., et al. 2001, A\&A, 365, L104
\bibitem[Rangarajan et al.(1995)]{rang95} Rangarajan, F. V. N., Fabian, A. C., Forman, W. R., \& Jones, C. 1995, MNRAS, 272, 665 (RFFJ)
\bibitem[Roettiger et al.(1996)]{roett96} Roettiger, K., Burns, J. O., \& Locken, C. 1996, ApJ, 473, 651
\bibitem[Saglia et al.(2000)]{saglia00} Saglia, R. P., Kronawitter, A., Gerhard, O., \& Bender, R. 2000, ApJ, 119, 153
\bibitem[Sarazin \& White(1987)]{Sar87} Sarazin, C. L., \& White, R. E. 1987, ApJ, 320, 32
\bibitem[Sarazin(1988)]{Sar88} Sarazin, C. L. 1988, "X-ray Emission from Clusters of Galaxies", ed. Cambridge University Press
\bibitem[Sarazin, Irwin \& Bregman(2000)]{Sar00} Sarazin, C. L., Irwin, J. A., \& Bregman, J. N. 2000, ApJ, 544, L101
\bibitem[Sarazin, Irwin \& Bregman(2001)]{Sar01} Sarazin, C. L., Irwin, J. A., \& Bregman, J. N. 2001, ApJ, 556, 533
\bibitem[Serlemitsos et al.(1993)]{ser93} Serlemitsos, P. J., Loewenstein, M., Mushotzky, R. F., Marshall, F. E., \& Petre, R. 1993, ApJ, 413, 518
\bibitem[Smith et al.(1996)]{Smith96} Smith, E. P., et al. 1996, ApJS, 104, 287
\bibitem[Snowden et al.(1994)]{Snow94} Snowden, S. L., McCammon, D., Burrows, D. N., \& Mendenhall, J. A. 1994, ApJ, 424, 714 (SMB)
\bibitem[Schombert(1986)]{schom86} Schombert, J. M. 1986, ApJS, 60, 603
\bibitem[Sulkanen \& Bregman(2001)]{Sulk01} Sulkanen, M. E., \& Bregman, J. N. 2001, AJ, 548, L131
\bibitem[Sun et al.(2000)]{sun00} Sun, M., Murray, S. S., Markevitch, M., \& Vikhlinin, A. 2000, to appear in the proceedings of the American Astronomical Society, HEAD meeting \#32, \#13.05
\bibitem[Tamura et al.(2001)]{Tam01} Tamura, T., et al. 2001, A\&A 365, L87  
\bibitem[Toniazzo \& Schindler(2001)]{toni01} Toniazzo, T., Schindler, S. 2001, MNRAS, 325, 509
\bibitem[Trinchieri \& Fabbiano(1985)]{Trin85} Trinchieri, G., \& Fabbiano, G. 1985, ApJ, 296, 447
\bibitem[White(1992)]{White92} White, D. A. 1992, PhD thesis, Univ. Cambridge
\end{thebibliography}
\end{document}